\documentclass[final,1p,times,twocolumn]{article}

\usepackage{amssymb}
\usepackage{amsmath}

\usepackage{graphicx}
\usepackage{dcolumn}
\usepackage{bm}
\usepackage{subcaption} 
\usepackage[dvipsnames]{xcolor}
\usepackage{cancel}
\usepackage[utf8]{inputenc}
\usepackage[T1]{fontenc}
\usepackage{mathptmx}
\usepackage{etoolbox}
\usepackage{float} 
\usepackage{ulem} 
\usepackage{parskip}
\usepackage{siunitx}
\usepackage{nicefrac}
\usepackage{upgreek}
\usepackage{xspace}
\usepackage{enumitem}
\usepackage{hyperref} 

\graphicspath{ {./figures/} }

\newcommand{\IV}{$IV$\!}
\newcommand{\DWC}{DW current\xspace}

\newcommand{\CR}[1]{\textcolor{blue}{#1}} 
\renewcommand{\CR}[1]{#1}

\newcommand{\secret}[1]{}

\begin{document}
\onecolumn
\begin{center}
\LARGE
\textbf{Demonstration of domain wall current in \\
		MgO-doped lithium niobate single crystals up to 400\,$^\circ$C}

\vspace{1.0cm}

\normalsize
\textbf{
Hendrik Wulfmeier\textsuperscript{1,2},
Uliana Yakhnevych\textsuperscript{1},
Cornelius Boekhoff\textsuperscript{1},
Allan Diima\textsuperscript{1},
Marlo Kunzner\textsuperscript{1}, \newline
Leonard M. Verhoff\textsuperscript{3},
Jonas Paul\textsuperscript{3},
Julius Ratzenberger\textsuperscript{4},
Elke Beyreuther\textsuperscript{4},
Joshua G\"ossel\textsuperscript{4}, \newline
Iuliia Kiseleva\textsuperscript{4},
Michael R\"using\textsuperscript{5},
Simone Sanna\textsuperscript{3},
Lukas M. Eng\textsuperscript{4,6},
Holger Fritze\textsuperscript{1,2}
}

\vspace{0.5cm}

\small
\textsuperscript{1} Institut f\"ur Energieforschung und Physikalische Technologien, Technische Universit\"at Clausthal, \\
Am Stollen 19 B, Goslar, 38640, Germany

\textsuperscript{2} Forschungszentrum Energiespeichertechnologien, Technische Universit\"at Clausthal, \\
Am Stollen 19 A, Goslar, 38640, Germany

\textsuperscript{3} Institut f\"ur Theoretische Physik and Center for Materials Research (ZfM/LaMa), Justus-Liebig-Universit\"at Gießen, \\
Heinrich-Buff-Ring 16, Gießen, 35392, Germany

\textsuperscript{4} Institut f\"ur Angewandte Physik, Technische Universit\"at Dresden, \\
N\"othnitzer Stra{\ss}e 61, Dresden, 01187, Germany

\textsuperscript{5} Integrierte Quantenoptik, Institut für Photonische Quantensysteme (PhoQS), Universit\"at Paderborn, \\
Warburger Stra{\ss}e 100, Paderborn, 33098, Germany

\textsuperscript{6} Dresden-W\"urzburg Cluster of Excellence - EXC 2147, Technische Universit\"at Dresden, \\
Dresden, 01062, Germany

\vspace{0.5cm}

\end{center}

\begin{abstract}
Conductive ferroelectric domain walls (DWs) represent a promising topical system for the development of nanoelectronic components and device sensors to be operational at elevated temperatures. DWs show very different properties as compared to their hosting bulk crystal, in particular with respect to the high local electrical conductivity. The objective of this work is to demonstrate DW conductivity up to temperatures as high as \SI{400}{\degreeCelsius} which extends previous studies significantly. Experimental investigation of the DW conductivity of charged, inclined DWs is performed using \SI{5}{\mole\percent} MgO-doped lithium niobate single crystals. \CR{Current-voltage (\IV) curves are determined by DC electrometer measurements and impedance spectroscopy and found to be identical. Moreover, impedance spectroscopy enables to recognize artifacts such as damaged electrodes. Temperature dependent measurements} over repeated heating cycles reveal two distinct thermal activation energies for a given DW, with the higher of the activation energies only measured at higher temperatures. Depending on the specific sample, the higher activation energy is found above \SI{160}{\degreeCelsius}~to~\SI{230}{\degreeCelsius}. This suggests, in turn, that more than one type of defect/polaron is involved, and that the dominant transport mechanism changes with increasing temperature. 
First principles atomistic modelling suggests that the conductivity of inclined domain walls cannot be solely explained by the formation of a 2D carrier gas and must be supported by hopping processes. This holds true even at temperatures as high as \SI{400}{\degreeCelsius}.
Our investigations underline the potential to extend \DWC based nanoelectronic and sensor applications even into the so-far unexplored temperature range up to \SI{400}{\degreeCelsius}.
\end{abstract}

\vspace{0.5cm}

\begin{center}
\end{center}

\vspace{3.5cm}

\Large
	\textbf{Research highlights:}

\normalsize	 
		\begin{itemize}
			\item First demonstration of high-temperature domain wall current.
			\item \CR{Current-voltage curves as determined by direct current electrometer measurements and impedance spectroscopy  are found to be identical.}
			\item The measured domain wall conductivity is traced back to hopping processes.
			\item Atomistic models demonstrate that domain wall segments parallel to the X and Y crystal directions remain insulating up to \SI{400}{\degreeCelsius}.
			\item \CR{Doping of lithium niobate with \SI{5}{\mole\percent} MgO has only a minor effect on the electronic structure at the domain walls.}
		\end{itemize}

\twocolumn
\normalsize
\section{Introduction}

Single-crystalline lithium niobate (LN) is a ferroelectric model material for applications based on piezoelectricity, electro-optical effects, or polarization switching. Devices include precision actuators, non-volatile ferroelectric random-access memories, electro-optical modulators, photovoltaic cells, resistive switches, and sensors for measuring vibration and magnetic fields, even at higher temperatures \cite{Poberaj2012,Waser2004}. 

One key factor that contributes to the versatility of LN-based devices is the ability to elegantly engineer ferroelectric domains and, thereby, domain walls (DWs) into the crystals \cite{Waser2004,Meier2021,McCluskey2022,Kirbus2019,Geng2021}. Domain engineering in LN involves the controlled creation and manipulation of DWs \cite{Shur2015}. Recently, considerable advancements have been reported for fabricating and tuning the DW properties in LN crystals, specifically also to reproducibly increase the DW conductivity by several orders of magnitude \cite{Kirbus2019,Werner2017,Godau2017,Zhang2022}. Application areas for customized DW structures are integrated optoelectronic components that extend the already proven use in the field of optical technologies, such as frequency doubling \cite{Feng1980}, with electronic functionality. Rectifiers \cite{Zhang2021} and a 2-terminal memory with very high memory window and long endurance \cite{Kaempfe2020} have already been demonstrated at room temperature. Further, manipulation and, in particular, enhancement of the \DWC has a broad impact for various devices, such as resistive switches, diode structures, or neuromorphic computing \cite{Zhang2021,Sharma2023,Chai2021, Suna2022}. 

For sensing devices, the ability to control the \DWC in LN offers completely new types of sensors when combined with specific functional films. In many cases they need to meet challenging environmental conditions such as elevated temperatures when used e.g. in aerospace applications or in the chemical industry \cite{Turner1994}. Hypothetical examples exploiting the strongly localized \DWC are sensors consisting of LN plates with gas-selective films (TiO\textsubscript{2}, SnO\textsubscript{2}, etc.), where the electrochemical potential of the latter and thus the sensor behavior is locally affected by the \DWC. Here, the DW provides the functionability of a VIA (Vertical Interconnect Access) thanks to its increased conductivity. Those devices must operate typically above \SI{200}{\degreeCelsius} to enable sufficiently fast reactions at the sensor surface.

The primary objective of this work is to demonstrate the \DWC in MgO-doped LN DWs at elevated temperatures up to \SI{400}{\degreeCelsius}. \CR{Thereby a potential effect of the dopant must be considered.} This allows us to prove the existence of thermally-activated processes beyond room temperature. If present, temperature ranges with certain activation energies should be identified. Further, to support whether or not the model of a 2-dimensional carrier gas accounts also for the DW current observed experimentally, density functional theory (DFT) calculations of the electronic structure of DWs in LN are presented.

Two different approaches to record the \IV\ characteristics of the DWs will be applied and compared to improve the reliability of the data:
\begin{enumerate}
	\item The direct determination of \IV\ characteristics using an electrometer, which involves direct current (DC) measurements; and 
	\item Alternating current (AC) measurements using a frequency response analyzer and analysis of the recorded impedance spectra. 
\end{enumerate} 

Each of these two methods has significant advantages and disadvantages, which should be discussed using the LN-DW measurement data, in particular with regard to the recognition of measurement artifacts.

\section{State of the art}\label{art}

LN is a ferroelectric with an extremely high Curie temperature of \SI{1160}{\degreeCelsius} and is a uniaxial ferroelectric, which supports only two opposing domain configurations along the z-axis, which are usually referred to as z$^+$ or z$^-$ depending on their orientation with respect to the surface. Lithium niobate typically comes in the form of large (up to \SI{8}{inch} in diamater), monodomain single-crystals, which make it ideal for domain engineering.

\begin{figure}[t]
	\centering
	\includegraphics[width=8cm]{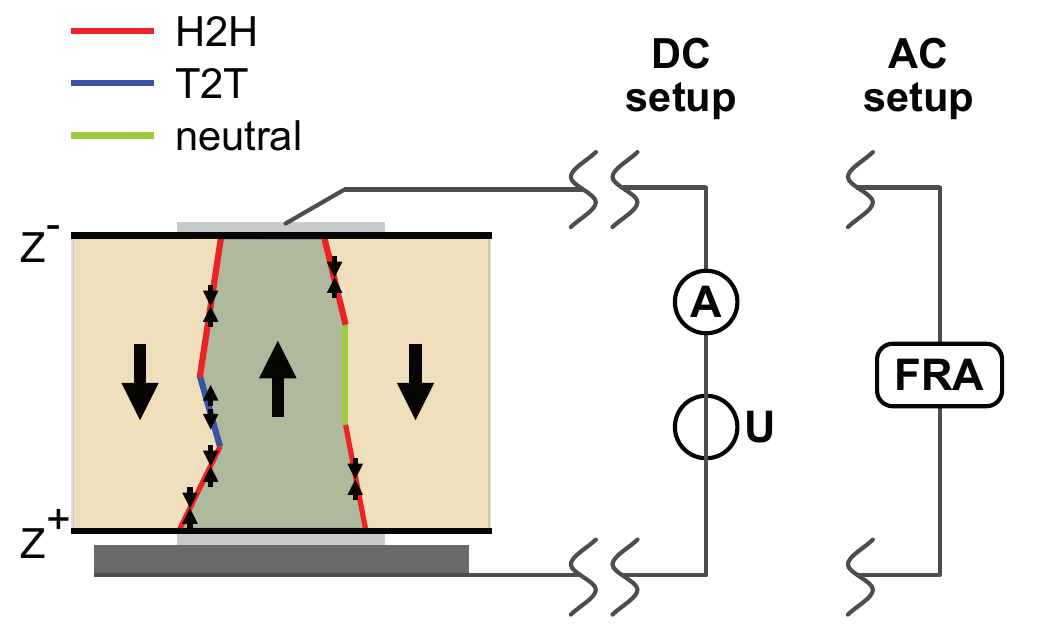}
	\caption{Schematic cross-section of a LN sample with enhanced predominantely head-to-head (H2H, red segments) and predominantely tail-to-tail (T2T, blue segments) domain walls (DWs). Electrical contacts and connectors are shown in light and dark gray, respectively. The two measurement configurations are shown on the right: DC setup using a precision electrometer and AC setup for impedance spectroscopy using a frequency response analyzer (FRA). The visualization concept for the DW structure is adapted from \cite{Godau2017}.}
	\label{DW-Schema}
\end{figure}

In general, real DWs do not extend as a uniform straight plane between the surfaces of a (single) crystal but consist of a large number of small segments with different inclinations, as it is depicted in Fig. \ref{DW-Schema}. A distinction is made according to the orientation of the local polarization vectors with respect to the domain wall. If the polarization vectors point towards each other (depicted as red DW segments in Fig.~\ref{DW-Schema}), this is referred to as a H2H-DW (head-to-head), and if they point in opposing directions (blue DW segment) this is referred to as a T2T-DW (tail-to-tail). The inclination is described by the angle $\upalpha$. If the domain wall runs parallel to the crystallographic z-axis, this is referred to as a n-DW (neutral DW). 
\CR{Due to the crystal's geometry, H2H and T2T DW segments can only have multiple lengths of the unit cell. For non-enhanced DWs, their width can be assumed to be few unit cells (see, e.g., Fig.~3c in \cite{Gonnissen2016}), which corresponds to the maximum length of the T2T and H2H sections. The enhancement leads to an inclination of the DW, which is accompanied by a slight change in the ratio of T2T and H2H sections, but hardly affects the DW width.}

Even at room temperature, the mechanism of charge transport along the DWs is the subject of ongoing research. In general, the inclination angle $\upalpha$ of the DW towards the polar axis has been identified theoretically \cite{Vul1973} and experimentally \cite{Godau2017} as the decisive parameter for the charge carrier density. The charge density at the DW increases with increasing $\vert\upalpha\vert$. 
Simulations underpin both the influence of $\upalpha$ and the poorer conductivity of T2T compared to H2H DWs, which correlates with experimental observations \cite{Eliseev2011,Beccard2023,Singh2022}. 
For their energetic stabilization, for example, a stronger shielding by electrons takes place in the case of an H2H wall, so that the charge carrier density in the immediate vicinity of the DW increases sharply.

Again, DWs in ferroelectrics with high spontaneous polarization exhibit properties that differ significantly from the volume of the corresponding single crystal \cite{Vul1973}. While it is established that only (macroscopically) inclined  DWs are conductive, no clear correlation between the inclination angle and \DWC is observed yet. Large conductivity variations are found even if all materials and DW preparation parameters are virtually identical \cite{Ratzenberger24}. DW current variations could be caused by minor differences during poling or intrinsic fluctuations of the materials properties within the crystal, including near-surface DW inclination, DW pinning and local heterogeneities of defects. As a result, the maximum \DWC among nominally identical samples can vary by several orders of magnitude \cite{Ratzenberger24}. 
Moreover, the crystal-electrode interface might contribute to the so far unexplained fluctuations in the charge transport \cite{Godau2017}. Note that the strongly varying \DWC among different samples mentioned above is not the focus of this high-temperature study of the DW current.

It is important to highlight that two distinct \IV\ characteristics of \DWC have been observed: ohmic and non-linear diode-like behavior. Specifically, non-linear diode-like behavior, proposed to be caused by space charges in the DW or at the DW-electrode interface, can be modeled by parallel resistor-diode pairs, leading to differing \IV\ characteristics due to the unequal properties of the z$^+$ and z$^-$ LN surfaces \cite{Zahn2024, Sanna17}.

Only limited DW current data is given in the literature in the vicinity of room temperature. Moreover, the authors are not aware of any studies at high temperatures at several hundred degree Celsius. In fact, despite some minor hints towards a thermally-activated \DWC process up to \SI{70}{\degreeCelsius}, studies on this topic have only rarely been reported \cite{Werner2017,Zahn2024} and the high-temperature behavior of domains and DWs remains unclear to date. 

By today, the electrical transport mechanism along charged ferroelectric DWs in LN and also in many other ferroelectrics is understood partly, only \cite{McCluskey2022,Zahn2024,Beccard2023,nataf20}. While the formation of a 2-dimensional electron gas (2DEG) at strongly charged walls is proposed and experimentally investigated in previous studies \cite{nataf20,bed18,bec22}, \DWC measurements also support a thermally-activated process at less-inclined and less-charged DWs. For example, in LN activation energies for the \DWC of about \SI{110}{\milli\electronvolt} and \SI{230}{\milli\electronvolt} are found \cite{Zahn2024}, however, \SI{100}{\milli\electronvolt} has also been reported \cite{Werner2017}. This contradicts the metallic-nature of a 2DEG, where no signatures of thermal activation should appear. Hence, different \DWC cases result: 
\begin{itemize}
	\item The band-bending below the Fermi-level due to strong electric field discontinuities at inclined and charged DWs, leading to the formation of a 2DEG \cite{nataf20,bed18}.
	\item \CR{Bandgap reduction due to an intrinsically different electronic band structure within the domain wall \cite{nataf20}.}
	\item Defect mediated charge carrier hopping processes (e.g., polaron hopping) along the DWs \cite{Beccard2023,xia18}. 
\end{itemize}

\begin{figure}[htb]
	
	\centering
	\includegraphics[width=8cm]{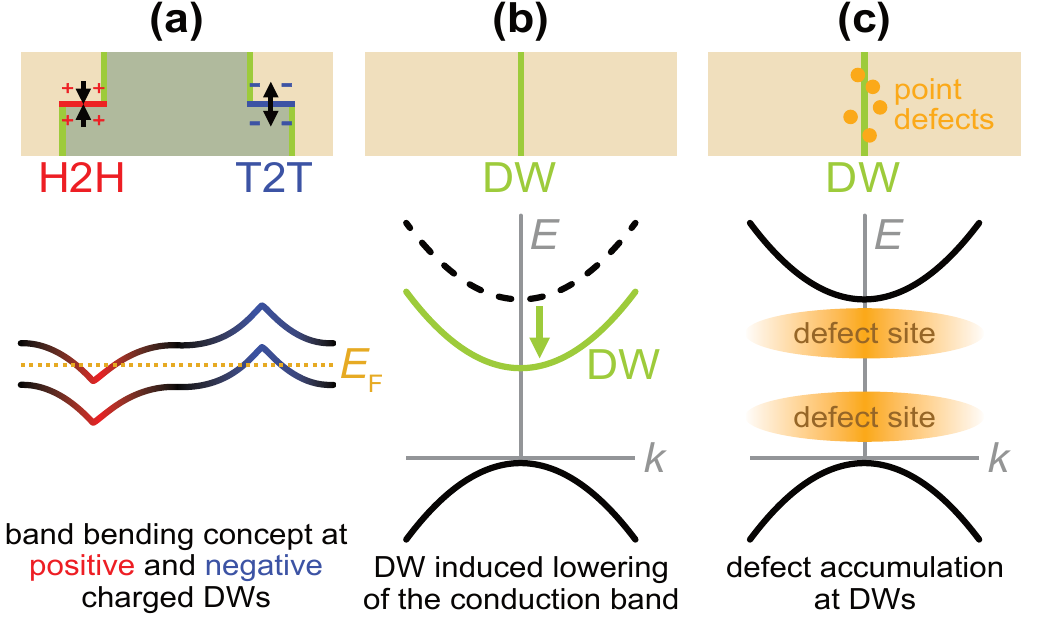}
	\caption{\CR{Potential DW effects on the band structure in lithium niobate (concepts adapted from  \cite{nataf20}). 
			(a)~Band bending results in states below or above the Fermi energy $E\textsubscript{F}$ due to positively and negatively charged H2H and T2T DW segments, respectively.
			(b)~Bandgap reduction due to an intrinsically different electronic band structure within the domain wall.
			(c)~Accumulation of point defects in or close to the DW. Defect sites generate additional energy levels within the bandgap.}}
	\label{Bandstruktur}
	
\end{figure}

\CR{These concepts are visualized in Fig.~\ref{Bandstruktur}.}

Based on spectroscopic investigations, it is assumed that strongly localized electron polarons (small polarons) are responsible for the charge transport in domains (bulk) as well as in DWs \cite{Reichenbach2018,Koppitz1987,Schirmer2009}. The relationship between \DWC $I$ and temperature $T$ can be expressed by:

\begin{equation}
	I(T) \propto \exp [-{E\textsubscript{\!A}}/{(k\textsubscript{B} T)}]
	\label{eq:EA}
\end{equation}

Here, $E\textsubscript{\!A}$ is the activation energy and $k\textsubscript{B}$ is the Boltzmann constant.

\subsection{\CR{MgO-doped LN}} 
\CR{MgO doping of LN results in threshold values of around \SI{3.0}{\mole\percent} and \SI{5.5}{\mole\percent} for the occurrence of different defects or defect complexes \cite{Sidorov2019}. These are associated with changes in the Li vacancy concentration. Specifically, below \SI{3.0}{\mole\percent} a decrease, from \SIrange{3.0}{8.0}{\mole\percent} an increase and above \SI{8.0}{\mole\percent} again a decrease in the Li vacancy concentration is observed \cite{iyi1995}. 
	\secret{Furthermore, it is expected that Mg incorporation at sites of $\text{Nb}_{\text{Li}}^{4+}$ polarons causes a reduction in the polaron concentration and thus in electron conduction. }
	At the MgO concentration relevant for this work, two Nb$_{\text{Li}}$ antisite defects are replaced by $\text{Mg}^{2+}$ and $\text{Mg}^{2+}$ starts to occupy $\text{Li}^{+}$ and $\text{Nb}^{5+}$ sites of the undisturbed structure by forming $\text{Mg}^{+}$ and $\text{Mg}^{3-}$ defects, respectively. Thereby a self-compensating pair $\text{Mg}^{+}$--$\text{Mg}^{3-}$ is created \cite{Sidorov2019} which does not affect the concentration of electronic charge carriers. Such a defect interacts with protons. However, their concentration is low as shown in \cite{Kofahl2024} for similar samples as long as no hydration treatment is done.
	As the DW current is carried by electron polarons, the effect of lithium vacancies or protons must not be taken into account. Most important, a modified polaron concentration is not expected due to the above mentioned self-compensating defect pair. Therefore, the question remains whether MgO doping has an effect on the electronic structure at the domain walls.}

\section{Experimental details}

\subsection{Samples}
In our experiments, z-cut \SI{5}{\mole\percent} MgO-doped LN wafers from Yamaju Ceramics Co., Ltd. (Japan), are used for domain engineering. 
Doping LN with MgO is motivated by an approximately \SI{50}{\percent} lower coercivity compared to undoped crystals \cite{Wengler2005}. \CR{This facilitates the targeted creation of domains with opposite polarization orientation during growth and DW pinning during poling in a monodomain crystal.}

Every sample piece measures \SI{200}{\micro\meter} in thickness, and has a lateral size of \mbox{$5\times\SI{6}{\square\milli\meter}$}. These samples were coated with electrodes (Cr thin films for room temperature and PtRh thin films for high-temperature characterization). For details on electrode deposition see \ref{app_electrodes}.

\subsection{Preparation of domain walls}
Single, hexagonally-shaped domains with a 'diameter' of \SIrange{110}{300}{\micro\meter} are fabricated by means of the UV-assisted electric field poling technique. For details see \cite{Godau2017,Haussmann2009} and \ref{app_DW-preparation}. 

In total, we investigated three different samples, tagged as LN1, LN2, and LN3. Note that samples LN1 and LN3 carry one such poled domain denoted by LN1-a and LN3-a, respectively, while two broadly separated domains were poled into sample LN2, as denoted LN2-a and LN2-b. Domain formation is monitored $in$-$situ$ via polarized light microscopy. The polarized light microscopy images illustrating the domains are shown in the inset of Fig.~\ref{fig:1} and in the \ref{app_domains}. 
The \DWC is enhanced following the protocol of Godau et al. \cite{Kirbus2019,Godau2017}. Details of this procedure are given in the \ref{app_enhancing}.

\subsection{\DWC measuring set-up} \label{experimental_setup}

The \DWC is investigated in the temperature range of \SIrange{25}{400}{\degreeCelsius} using a high-temperature micro-impedance setup which basically consists of a heating module and  electrical micro-probes where an impedance spectrometer or an electrometer is connected (see below). The samples are placed atop the heater. Details of the setup can be found in the \ref{app_micro-impedance-setup}.

Individual domains are contacted by a platinum tip that is precisely positioned using a 3D micromanipulator. The microimpedance system used here allows precise positioning of measuring tips on the samples so that both the direct current and alternating current resistance (impedance) of samples with small electrode surfaces can be determined. 

In order to compare the results of the DC voltage measurements and the impedance spectroscopy, these are determined on the same samples with the identical wiring and in temporal nearness. The applied measurement voltage $U\textsubscript{M}$ is provided by the respective devices.
The temperature was varied from room temperature up to \SI{400}{\degreeCelsius} when determining the \IV\ characteristics. The AC measurements were generally carried out at constant temperature levels, whereby the system was allowed to relax for at least  \SIrange{10}{20}{\min} before each measurement was started. DC electrometer data were continuously recorded during the constant temperature ramps ranging from \SIrange{0.5}{3.5}{\kelvin\per\min}. In between, these were shortly paused each approximately \SI{20}{\kelvin} for a few minutes in order to perform \IV\ curve recordings. For all measurements, the total pressure in the measuring chamber was lowered to < \SI{2}{\pascal}. The wiring was chosen so that the variable voltage is applied to z$^+$. The z$^-$ electrode is at earth potential (see Fig.~\ref{DW-Schema}).

Direct current \IV\,sweeps between $\pm$\SI{10}{\volt} are recorded using a precision electrometer (Keithley 6517B, US). The step widths vary from \SI{2}{\volt} (for overview measurements) to \SI{0.05}{\volt}. The recorded measured values were formed from the average of 5 individual measurements each with an integration time of \SI{200}{\milli\second}.

For the \DWC characterization with alternating current (AC) an impedance spectrometer (Solartron SI 1260, UK) was used. In order to be able to measure sufficiently small currents, especially at low temperatures and/or low measuring frequencies,  the setup was extended by a dielectric high impedance interface (Solartron SI 1296, UK), which widens the measuring range up to about \SI{10}{\giga\ohm}. 
As with the determination of \IV\ characteristics, a DC bias voltage of $\pm$10\,V was applied. To measure the impedance, it is superimposed by an alternating voltage with an amplitude of \SI{50}{\milli\volt}\CR{, a value that is typically used in imepance spectroscopy. This approach is the precondition for measuring the impedance at the given charge carrier concentrations under bias~(see also \ref{app_enhancing}}.
Impedance spectra were recorded at a frequency range from \SI{1}{\mega\hertz} down to \SI{50}{\hertz} at a typical integration time of \SI{1}{\second}. 

The resistance is then extracted by fitting an equivalent circuit to the impedance spectra. It consists of a parallel connection of DW resistance $R\textsubscript{DW}$ and constant phase element ($R\textsubscript{DW}\!\parallel$\,CPE). Additionally, the resistance of the supply lines $R\textsubscript{supply}$ is connected in series to these. The CPE element is chosen instead of a pure capacitance in order to take capacity distributions within the DWs, inhomogeneities (roughness, plane parallelism, crystal defects, etc.) and finite expansion of the samples into account \cite{Hirschorn2010, Macdonald2005}. In some cases a second semicircle appears in the Nyquist diagram. Then, the equivalent circuit diagram is extended by an additional $R\! \parallel$\,CPE element connected in series.

\section{Results and discussion}

\subsection{Temperature-dependent domain wall IV characteristics} \label{T-depIV}

\subsubsection{Linear behavior}

\begin{figure}[htbp]
	\centering
	\includegraphics[width=8cm]{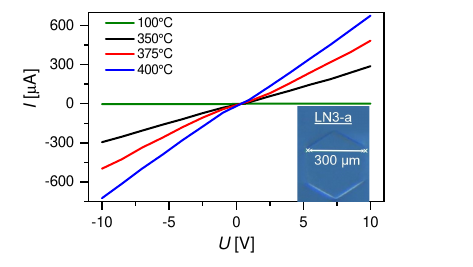}
	\caption{\IV\ characteristic across the $\pm$10~V voltage bias regime of the domain wall current in domain LN3-a. The inset depicts the DW contour of domain LN3-a.} 
	\label{fig:1}
\end{figure}

As-grown DWs show a negligibly small \DWC across the full $\pm$\SI{10}{\volt} range. Typically, values from \SIrange{10E-12}{10E-10}{\ampere} are found at room temperature (RT) at a bias voltage of +7\,V, see the \ref{app_domains}. The \DWC is enhanced by several orders of magnitude as shown exemplary by the \IV\ characteristics in Fig.~\ref{fig:1} for domain LN3-a at different temperatures. Remarkably, the \DWC at \SI{400}{\degreeCelsius} is increased at least 800 times as compared to the RT \DWC at the bias voltage mentioned above. Additionally, \ref{app5} provides \IV\ curves, showing similar increases in the \DWC for all studied domains across various temperatures.

Despite nominally identical preparation of the domains and all pieces originating from the same 75\,mm wafer, the \IV\ curves show significant variations in magnitude. The statement applies for DW currents both before and after \DWC enhancement. These differences already mentioned above are known from literature and may be due to minor differences in sample preparation \cite{Godau2017,Ratzenberger24}. 
An \IV\ characteristic as shown in Fig.~\ref{fig:1} is an example for the nearly linear case.

\CR{Note that DW currents are given instead of conductivities, since the DW cross-section as required to calculate the conductivity is not known or at least stongly dependent on the definition of the DW width. According to \cite{Eliseev2011}, the overall width of a DW is estimated to be in a range from \SIrange{20}{100}{\nano\meter} whereas the holes accumulate mainly in a region of \SIrange{2.5}{5}{\nano\meter}. As electrons are the dominant charge carriers dictating the DW current, the effective DW width for DW conductivity is quite likely on the order of several \SI{10}{\nano\meter} which could further depend on an unknown local roughness as indicated by TEM measurements \cite{Gonnissen2016}. At room temperature DW currents range from \SIrange{E-7}{E-6}{\ampere} at a +9\,V bias. The observation goes along with the statement in section~\ref{art}, that large variations of the DW current occur even when all materials and DW preparation parameters are kept identical. Notably, the DW current studied here shows no significant dependence on the DW length (domain circumference) presumably due to the overwhelming effect of varying DW inclination angles.}

\CR{A rough estimate of the DW conductivity can be obtained considering the minimal and maximal DW widths
	of \SI{2.5}{\nano\meter} and \SI{100}{\nano\meter}, respectively, given in Ref.~\cite{Eliseev2011}. With a typical DW circumference of our samples of $\sim$\SI{1000}{\micro\meter}, DW cross-sections of
	\SI{2.5}{\square\micro\meter} and \SI{100}{\square\micro\meter} result, respectively. For a current of
	\SI{500}{\micro\ampere} as found roughly at \SI{400}{\degreeCelsius} and \SI{9}{\volt} (compare
	Fig.~\ref{fig:1}), a DW conductivity between \SI{45}{\siemens\per\centi\meter} and
	\SI{1}{\siemens\per\centi\meter} follows, respectively, assuming a uniform sheet resistance for that DW.
	Those values are several orders of magnitude larger than reported for (undoped) lithium niobate
	($\sim$\SI{E-8}{\siemens\per\centi\meter}) at the same temperature \cite{Kofahl2024}.}

\subsubsection{\CR{Non-linear behavior}} \label{DC_diode}

\begin{figure}[t]
	\centering
	\includegraphics[width=8cm]{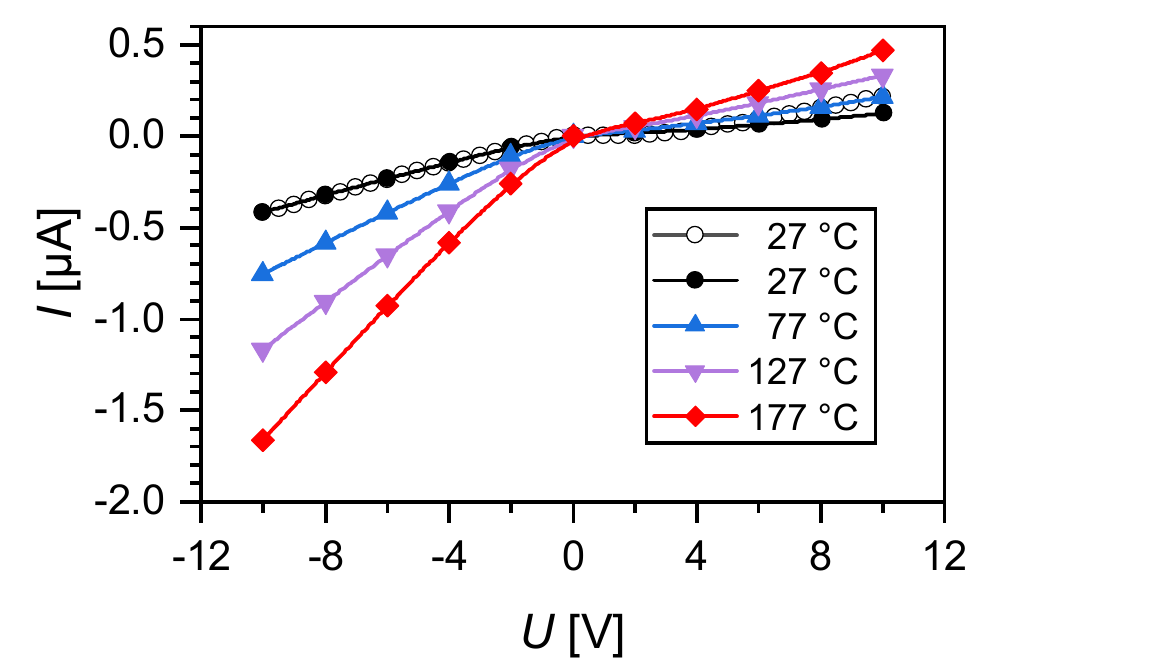}
	\caption{\IV\ characteristic of LN1-a at selected temperatures characterized with DC measurements. A comparative measurement about 9 months earlier at 300 K is shown with circles. The almost identical curve suggests that no significant degradation (such as relaxation of the DWs) took place during the previous experiments below room temperature \cite{Zahn2024}.}
	\label{DC-IV}
\end{figure}

Fig.\,\ref{DC-IV} shows the \IV\ characteristics of the DW LN1-a resulting from the direct current (DC) measurements for different temperatures. The measured DW currents are in the order of \SI{1}{\micro\ampere}. Room temperature measurements on similar samples without a DW structure yield currents <\,\SI{1}{\pico\ampere} and are therefore approx. 6 orders of magnitude below the DW currents found here (compare the following section\,\ref{sec:Temp-dep_DWC}).
The DW characteristics show an asymmetrical behavior. For negative bias voltages, a linear, almost ohmic behavior results. Positive voltages result in a non-linear partial characteristic similar to that of a diode. 
Higher currents can be measured as the temperature rises. In the temperature window from room temperature to \SI{175}{\degreeCelsius}, the current increases by approximately factor 4. 

In previous works \cite{Kirbus2019,Godau2017,Zahn2024} often rectifying characteristics have been reported as well. This bahavior can be assigned to Schottky barriers at the electrode-DW junction \cite{Zahn2024}. Given the numerous local and nanoscopic influences in DW preparation, the presence of Schottky barriers can hardly be controlled. 

\subsection{AC impedance spectroscopy}
\begin{figure}[htbp]
	\centering
	\includegraphics[width=8cm]{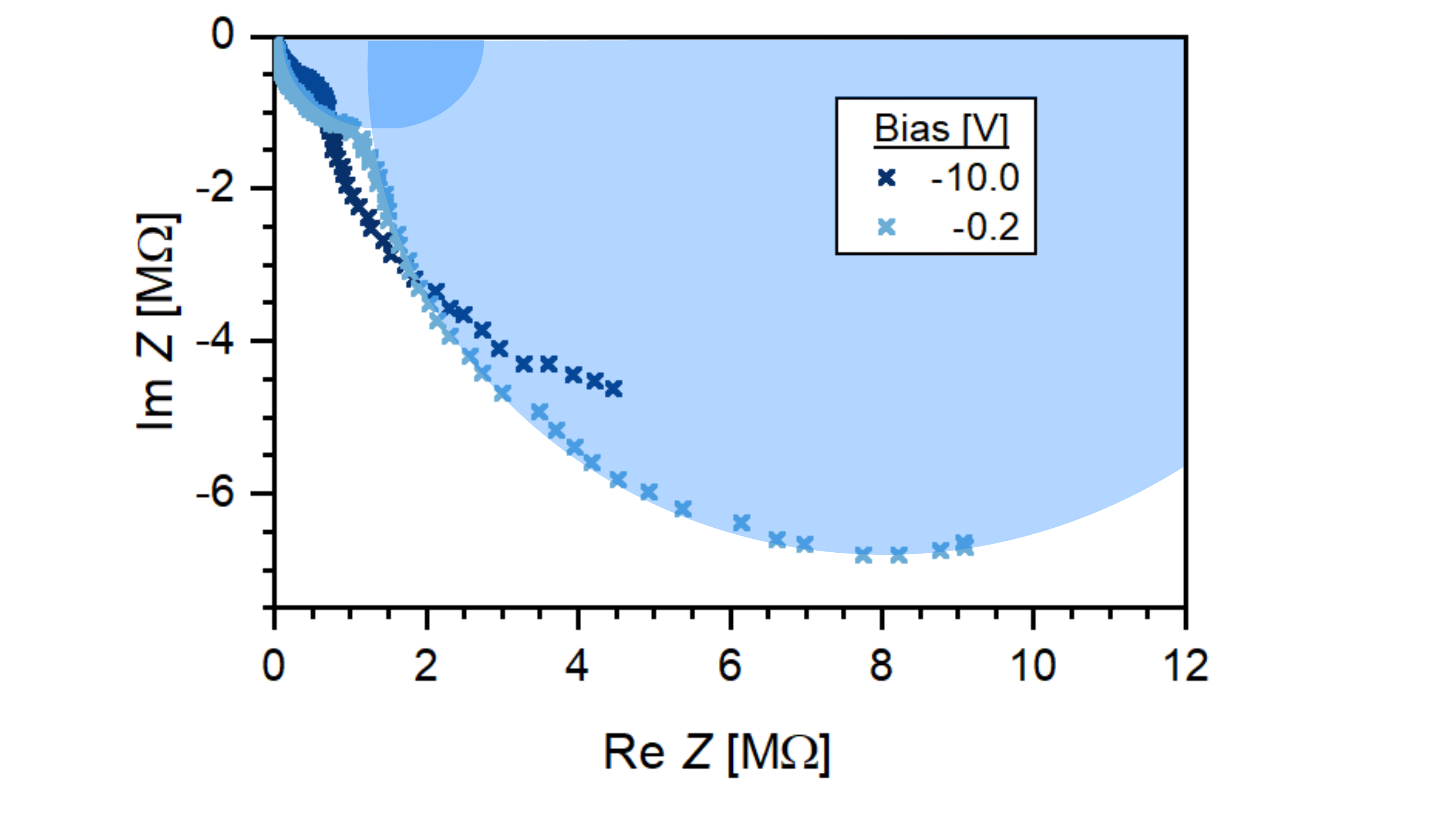}  
	\caption{DW impedance spectra with different bias voltages, recorded at \SI{100}{\degreeCelsius} during the first measurement cycle of LN1-a. For the sake of clarity, only two exemplary data sets are shown.}
	\label{AC-IV-2SC}
\end{figure}

\begin{figure}[htbp]
	\centering
	\includegraphics[width=8cm]{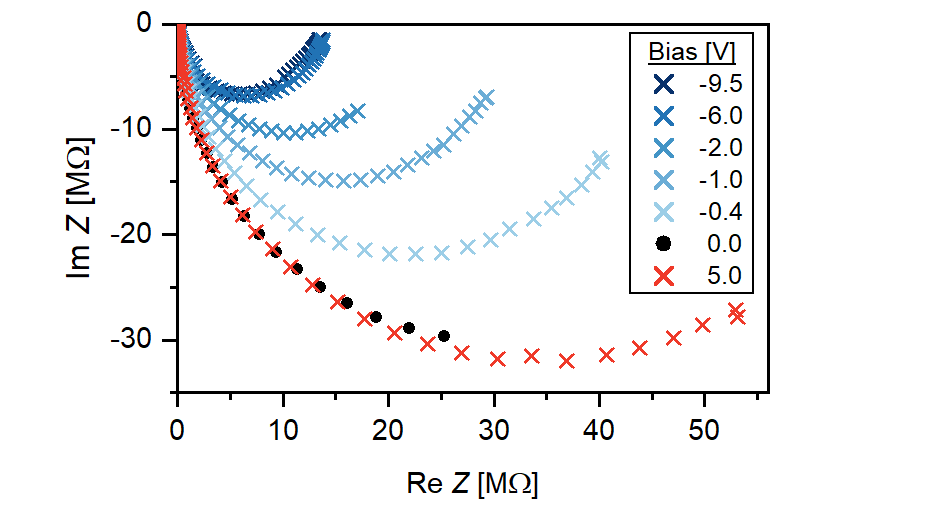}
	\caption{Repeated measurement of the DW impedance spectra at \SI{100}{\degreeCelsius}. The partially degraded electrodes were previously regenerated by applying a layer of silver conductive paint.}
	\label{AC-IV-1SC}
\end{figure}

In addition to the DC measurements, impedance spectroscopy is carried out to investigate the DW behavior and potential effects of the electrodes in more detail.
Fig.~\ref{AC-IV-2SC} shows impedance spectra at different bias voltages in a Nyquist representation. Two incomplete, partially overlapping semicircles are visible for each voltage. The spectra were accordingly fitted with a series resistor and \underline{two} $R\!\parallel$\,CPE elements. 
The resistances resulting from the low-frequency semicircle match well with the 
data from the DC measurements. The smaller, high-frequency semicircle is approximately one order of magnitude smaller and cannot be meaningfully assigned to an expected behavior in relation to either a domain (bulk) or DW transport in LN. A measurement artifact can therefore not be ruled out. 
After completing the measurement cycle, degradation of the electrodes became visible, so that these were renewed by means of a silver conductive coating on the z$^+$ side. The measurement series was then repeated (see Fig. \ref{AC-IV-1SC}), whereby the additional high-frequency semicircle no longer occurred, so that a fit model with only one $R\!\parallel$\,CPE element was selected.
The real part of the low-frequency semicircles in Fig.~\ref{AC-IV-2SC} agrees well with the real part of the semicircles in Fig.~\ref{AC-IV-1SC}. At the same time, the capacitance values in the remaining semicircle increased by about \SI{10}{\percent}, which can be explained by and corresponds well with the increase in the electrode area due to the additional silver conductive varnish, which slightly increased the original electrode area of about the same factor. 

\CR{Furthermore, the capacity was determined from the CPE fit parameters according to \cite{Macdonald2005}, ranging from \SIrange{21.5}{22.1}{\pico\farad}, depending on the sample, e.g. the electrode area. This results in permittivities along the z-axis from 38.6 to 39.7 for the MgO-doped samples discussed in this work, respectively. Literature values for undoped LN single crystals range from 23.0 to 40.0 
	\cite{Kovacs1990,Kushibiki1999,Nakagawa1973,Smith1971,Warner1967,Xue2002,Yamada1967,Nassau1966,Mansingh1985,Teague1975,Morita2001}. 
	This work's values fall in the range of the literature data. Please note that the measured capacitances have not been corrected for stray capacitances which would slightly lower our experimental values.}

Two conclusions can be drawn by this. First, the correlation of the capacitance with the electrode area supports the hypothesis that the capacitance of the $R\!\parallel$\,CPE parallel circuit corresponds to the bulk properties, i.e.\ the domain, while the resistance reflects the conductivity properties in the DW\CR{, therefore, referred to as $R_{\text{DW}}$ and $C_{\text{bulk}}$ in the following}. Secondly, the disappearance of the additional semicircle \CR{($R\textsubscript{artifact}\!\parallel$\,CPE\textsubscript{artifact})} during the second measurement cycle is also an indication that the high-frequency semicircle in Fig.~\ref{AC-IV-2SC} is an artifact obviously due to poor contacting as a result of a degraded electrode. A notable change/degradation of the DW structure can therefore be ruled out as the cause with a high degree of probability.

\subsection{Comparison of DW IV characteristics between DC and AC measurements}
The advantage of the DC measurement method is clearly its high measurement speed. Even with a step width of only \SI{0.2}{\volt} and typical data acquisition speeds of \SI{0.7}{\second} to \SI{0.8}{\second} per data point,
it only takes approx.\,\SI{80}{\second} to record such a DW \IV\ characteristic curve. The measurement uncertainty in dependence of the respective data range is given in \ref{app_micro-impedance-setup}.
It should be noted that the measurements not only reflect the transport in the DWs, but can also be influenced by contact resistances which is one of the potential measurement artifacts that can be overseen easily.

Classical AC impedance spectroscopy using a frequency response analyzer can reflect individual contributions to the current and enables, therefore, a deeper understanding. However, the method records measured values significantly slower compared to direct current measurements. Depending on the selected measuring frequency interval (see section \ref{experimental_setup}), it usually takes \SIrange{3}{5}{\min} just to record an impedance spectra which results in a single data point only. 

An important question is the consistency between the DC electrometer measurements and the AC impedance spectroscopy on the same DW. To verify this, the current $I$ is calculated from the $R$ for a given voltage $U$ following Ohm's law $I=U\!/R$. 
The current in the AC impedance measurements can be reconstructed using the following integration:
\begin{eqnarray}
	I(U) =  \int\limits_{-10}^{U} \frac{1}{R(U)} \ \mathrm{d}U - I_0  \\
	\text{with the condition} \hspace{10pt} I_0 = 0 \hspace{10pt}\text{at}\hspace{10pt}  U = 0 \nonumber
\end{eqnarray}

Here, the integration constant $I\textsubscript{0}$ is chosen in a way that ensures the current being set to zero when the applied voltage is zero.

\begin{figure}[t]
	\centering
	\includegraphics[width=8cm]{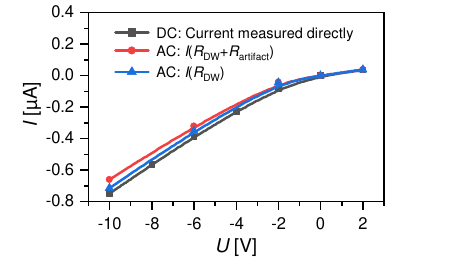}
	\caption{Comparison of the DW \IV\ characteristic curves recorded by electrometer (DC measurement) and impedance spectrometer (AC measurement) at RT. The \DWC of the AC voltage measurements was extracted from the impedance spectra by integrating the reciprocal resistances.}
	\label{AC-DC}
\end{figure}

To demonstrate the influence of the artifacts, i.e. the electrode degradation discussed above, on the \IV\ characteristics the data from the first measurement cycle (Fig.\,\ref{AC-IV-2SC}) were further evaluated in two ways. 
In the first case, the resistances of the two semicircles \CR{($R\textsubscript{DW}+R\textsubscript{artifact}$)} were added before integration. 
In the second case, the resistance \CR{$R\textsubscript{DW}$} of the low-frequency semicircle only was used to calculate the current according to Eqn.~2. \\
In Fig.\,\ref{AC-DC}, the resulting DW currents are plotted as \IV\ characteristic curves together with the DC current measured by the electrometer (see section \ref{DC_diode}). 
As errors accumulate during the integration, the initial range of the integration (negative voltages) is particularly meaningful. A comparison of the curves shows only minor differences between the individual measurement methods. Only the first AC measurement \CR{($I(R\textsubscript{DW}+R\textsubscript{artifact})$)} with the second semicircle, caused by electrode degeneration, diverges slightly from the other two characteristic curves. Most important, such an artifact is clearly visible as a further semicircle in a Nyquist representation which motivates the application of AC impedance spectroscopy as background method.

\subsection{\label{sec:Temp-dep_DWC}Temperature dependent domain wall current}

\begin{figure}[htbp]
	\centering
	\includegraphics[width=8cm]{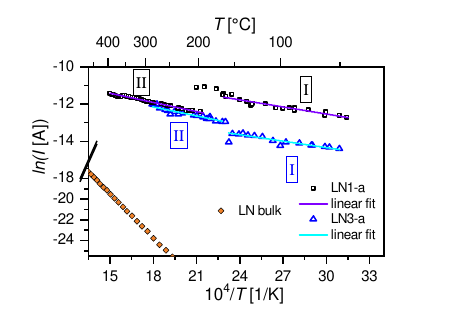}
	\caption{Temperature-dependent \DWC plotted for domain LN1-a (\textcolor{purple}{purple}) and LN3-a (\textcolor{blue}{blue}) at a +7~V bias voltage. For reference, the bulk LN current (in \textcolor{orange}{orange}) is displayed in the same Arrhenius plot.}
	\label{fig:2}
\end{figure}

Fig.~\ref{fig:2} displays the \DWC recorded at a $+$7~V bias voltage for domains LN1-a and LN3-a in the Arrhenius-type plot. The \DWC increases as a function of temperature for all domains (see Fig.~\ref{fig:2} as well as Figs. \ref{fig:A8} and F\ref{fig:A9} in \ref{app6}). This trend is observed consistently for repeated heating. 

Of particular interest is the comparison between investigated DW currents and the bulk LN current. The latter follows from conductivity data of a sample without DWs obtained through impedance spectroscopy \cite{Kofahl2024} and is calculated using the thickness and average electrode size of the samples with DWs. Note the significantly lower current magnitude of bulk LN (\textcolor{orange}{orange} data points) with respect to domains LN1-a and LN3-a shown in Fig.~\ref{fig:2}. In particular, a six-order-of-magnitude difference is found at \SI{250}{\degreeCelsius}. Therefore, we can exclude any noticeable bulk related contribution to the measured DW currents. 

Moreover, we notice in Fig.~\ref{fig:2} different temperature regimes that show a linear increase in \DWC in the Arrhenius-type representation, however, with different slopes, indicating thermally-activated transport processes with different activation energies. Further, DW currents for LN1-a and LN3-a differ by about two orders of magnitude near RT. This trend persists up to approximately \SI{200}{\degreeCelsius}, hence branding the Regime~I in Fig.~\ref{fig:2}. Above about \SI{200}{\degreeCelsius}, Regime~II starts, where \DWC data scatter less, and the differences gradually diminish, converging practically to the same \DWC value for all analyzed DWs. These findings suggest that the exact DW morphology plays no major role anymore for the given domains above about \SI{200}{\degreeCelsius}. Note that a closer view to all data delineates the temperature ranging into two overlapping regimes: room to medium temperatures (\SIrange{25}{230}{\degreeCelsius}) and temperatures above about \SI{160}{\degreeCelsius}. The transition temperature between the regimes depends obviously on the specific domain, which is not unexpected in view of the scattered DW properties as discussed above for the absolute \DWC.

\subsection{\label{sec:res_EA}Activation energy and polaron hopping processes}

\begin{figure}[htbp]
	\centering
	\includegraphics[width=8cm]{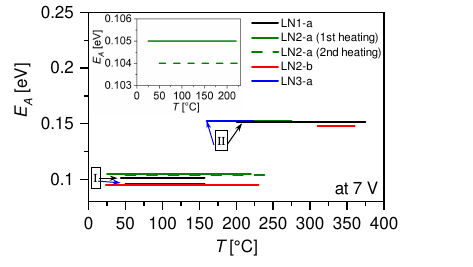}
	\caption{Activation energies $E_A$ of the \DWC for the distinct temperature ranges as marked in Fig.~\ref{fig:2}. Note the reproducible $E\textsubscript{A}$ values of DW LN2-a after two heating cycles.}
	\label{fig:3}
\end{figure}

Fig.~\ref{fig:3} presents the activation energy ($E\textsubscript{A}$) for all investigated DW currents as extracted by fitting Eq.\,\ref{eq:EA} to the linear Regimes I and II in Fig.~\ref{fig:2}. The uncertainty for the extracted $E_A$ values ranges up to $\pm 20$ meV for the four investigated DW currents.
The temperature intervals used for data fitting are the same ones as the ones delineated in Fig.~\ref{fig:2}. Fig.~\ref{fig:A10} in the \ref{app6} displays the activation energies across different temperature ranges as a function of time to illustrate the measurement sequence. In addition, Fig.~\ref{fig:3} contains also the as-deduced $E\textsubscript{A}$ for DW LN2-a, depicted for both its initial \DWC state (\textcolor{green}{green} solid line) and the subsequent heating cycle (\textcolor{green}{green} dashed line). Remarkably, our findings suggest that the studied DWs exhibit consistent behavior, with two distinct activation energy values of about $E\textsubscript{A}$\,=\,\SI{100}{\milli\electronvolt} and \SI{160}{\milli\electronvolt} that remain unchanged over broad temperature ranges if the measurement uncertainty is considered.

Werner et al. \cite{Werner2017} and Zahn et al. \cite{Zahn2024} reported $E_A$ values of \SIrange[range-phrase = --]{100}{110}{\milli\electronvolt} and \SI{230}{\milli\electronvolt} for the temperature range from \SI{-193}{\degreeCelsius} to \SI{70}{\degreeCelsius}. They also suggested that the dominant charge carrier type for \DWC in these DWs is of electronic nature, backed up by Hall-transport measurements \cite{Beccard2023} and strain-coupling experiments \cite{Singh2022}, specifically involving thermally activated electron-polaron hopping. It is worth noting that the activation energies of \SI{100}{\milli\electronvolt} and \SI{200}{\milli\electronvolt} were observed in two distinct samples, with each sample exhibiting one of these values. The authors explained this variation by differences in the distribution of electronic defect states at the metal-DW interface, introduced during the metal-electrode deposition and conductivity enhancement procedures. The same arguments can be chosen to explain the difference to the upper activation energy we found. Further, we assume that the activation energy values obtained here, also indicate involvement of electron-polaron hopping processes; in fact, such $E_A$ values align well with the reported values for electron-polaron jumps in bulk lithium niobate \cite{Reichenbach2018, Koppitz1987, Schirmer2009}. 	

It is evident from our experiments that DWs exhibit distinct activation energies across different temperature ranges. This implies the involvement of more than one type of defect/polaron that contributes to the \DWC.

\subsection{DFT modelling}\label{DFT}

\begin{figure}[htbp]
	\centering
	\includegraphics[width=8cm]{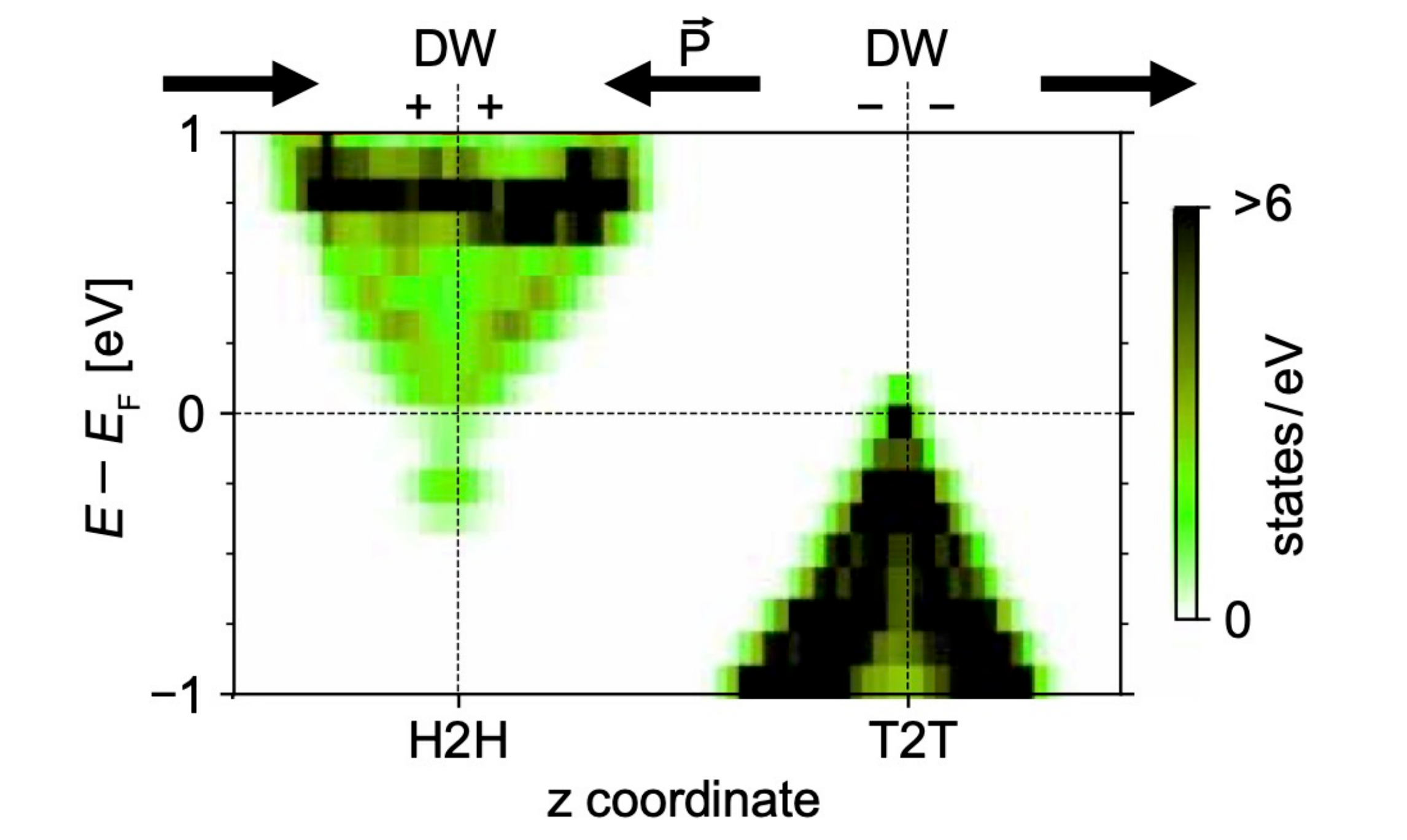}
	\caption{Local density of states (DOS) of a supercell modelling a fully-charged H2H and a T2T DW in LN. The direction of polarization, the position of the Fermi energy, and the sign of the polarization charges are indicated.}
	\label{fig:4}
\end{figure}

In order to investigate the mechanisms behind the \DWC, DWs perpendicular to the crystallographic directions X, Y and Z of stoichiometric LN are modeled within DFT. We thereby employ the VASP software package \cite{Kresse1996,Kresse1996_2}, projector augmented-wave (PAW) potentials implementing the Perdew-Burke-Ernzerhof (PBE) formulation of the xc-functional \cite{Perdew1996,Perdew2008,Bloechl94} and a cutoff energy for the plane wave basis of 500~eV. Further computational details can be found in the \ref{app_atomistic}. Structural optimization reveals that the atomic positions and the local polarization reach the value of bulk LN at a distance of about 1\,nm from the DW center in the case of X and Y DWs, and at a distance of 2\,nm in the case of the fully-charged Z walls. This is in agreement with typical DW widths as deduced from transmission electron microscopy investigations \cite{Gonnissen2016}. The presence of DWs deeply modifies the electronic structure  of the material's bulk, as shown by the calculated density of states (DOS) shown in the \ref{app_atomistic} (Fig.~\ref{fig:11}). 
However, while X and Y DWs reduce the fundamental bandgap from \SI{3.4}{\electronvolt} to about \SI{2.4}{\electronvolt} and \SI{3.0}{\electronvolt}, respectively, charged Z DWs lead to an overlap of valence and conduction band, as demonstrated by the local density of states of the super-cell modelling the DWs shown in Fig.~\ref{fig:4}. The direction of polarization and the sign of the polarization charges for the head-to-head (H2H) and the tail-to-tail (T2T) DWs are indicated as well. Interestingly,
the spontaneous polarization causes a strong band bending for both types of fully-charged DWs, so that a non-vanishing portion of electronic states belonging to the conduction (valence) band, crosses the Fermi energy in proximity to the H2H (T2T) DW \cite{Verhoff2024}. The domain walls thus become semi-metallic, similar to suggestions in other materials \cite{nataf20,bed18}. Moreover, the DOS at the H2H and T2T DWs is very different; H2H DWs feature very low DOS values at the Fermi-energy and higher DOS values about \SI{150}{\milli\electronvolt} below it. Unfortunately, due to the limited precision of the computational approach, it cannot be conclusively stated whether this gap in the H2H DWs DOS is related to the activation energy reported below or not.

According to our calculations, a 2DEG is formed only for fully-charged H2H and T2T DWs, that might contribute to the \DWC. However, inclined DWs, such as the DW measured in the present experiments, can be considered as a sequence of non-charged X and Y DWs and charged Z DWs (H2H and T2T) walls, where sections with the lower conductivity govern the overall behavior. This picture is supported by TEM Investigations, showing steps in the domain wall at the unit cell level \cite{Gonnissen2016}. Thus, inclined DWs will feature insulating segments, and the presence of H2H and T2T walls within an oblique DW will not suffice to render the whole DW metallic.

Addressing the interpretation of our DFT calculations in relation to the experiment, three different aspects must be considered. 

\begin{itemize}
	\item The calculations presented here model a \SI{0}{\kelvin} system. Vibrational contributions are expected to close the electronic bandgap as a function of the temperature. In a previous work, it was shown by \textit{ab initio} molecular dynamics and Allen-Heine-Cardona theory that temperature effects reduce the fundamental bandgap of LN by less than 1 eV at \SI{400}{\degreeCelsius} \cite{Riefer16}. Thus, X and Y DWs will remain insulating (with a bandgap of about \SI{2}{\electronvolt}) even considering temperature effects.
	\item Improved methods beyond DFT with (semi)local xc functionals as used here for calculating the electronic structure, are required to quantitatively estimate the overlap between conduction and valence bands in charged DWs. However, the DFT estimated bandgap of X and Y DWs represents a lower bound, and more refined calculations will not modify the conclusion that those DW segments are insulating.
	\item The role of high doping concentrations or defects, such as lithium vacancies is still unclear. They might easily lead to the formation of additional energy levels within the DWs (e.g., donor or acceptor levels) or shifts in the band structure. 
	\CR{The negligible influence of MgO doping as used in the current experiment on the above results is discussed in the following subsection.}
\end{itemize}

This calculation demonstrates the feasibility to describe DW band structures via first principle calculations and opens up the option of high DW currents. However, to which extent different sections and mechanisms such as the 2DEG formation, any polaron accumulation and charge hopping contribute to the \DWC remains to be settled with further investigations which must include the structure of the inclined DWs.

\subsubsection{\CR{DFT calculations on MgO-doped LN}}

\begin{figure}[htbp]
	\centering
	\includegraphics[width=8cm]{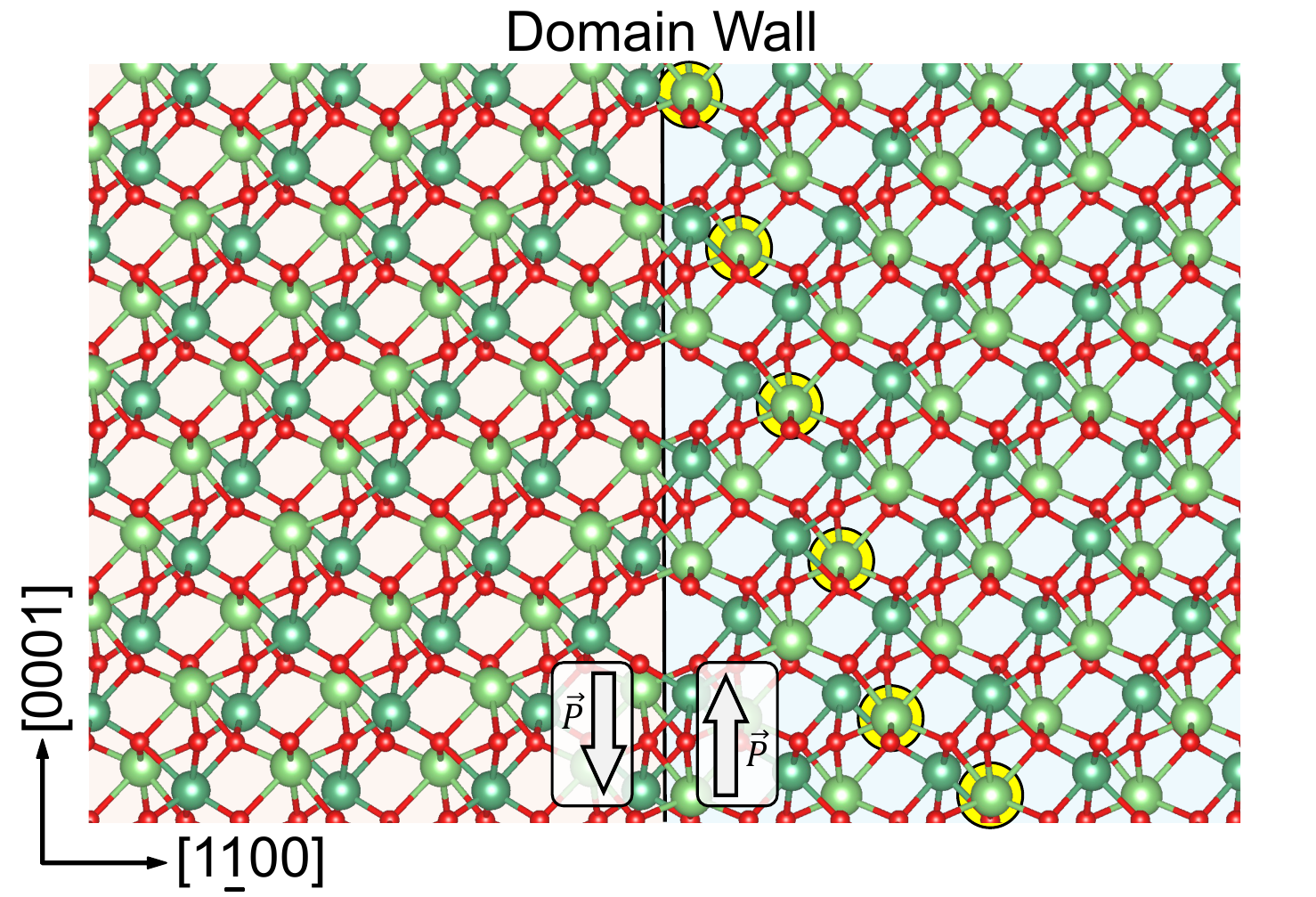}
	\caption{\CR{Atomic structure around a 180$^\circ$ X-DW in LiNbO$_3$ according to Ref. \cite{Verhoff2024}.
			The dashed line separates the ferroelectric domains with spontaneous polarization directed downwards (lhs)
			and upwards (rhs). Li and Nb atoms are represented in light and dark green, respectively. O atoms are in red,
			while the position at which the Mg substitutionals have been modelled are shown in yellow.}} 
	\label{crystalstructure}
\end{figure}

\CR{In order to explore the role of MgO doping on the electronic structure of the non-conducting
	X-DWs and Y-DWs, a doping concentration of up to \SI{2}{\mole\percent} has been furthermore
	considered in the 
	atomistic models. According to the literature, we consider Mg$_\text{Li}^{2+}$ substitutionals,
	which we model as a function of their distance from the domain wall. 
	Fig.~\ref{crystalstructure} depicts exemplarily the atomic structure of a X-type LN domain
	wall as calculated within DFT with the approach described in the previous section. On the right
	hand side of the structure the position are marked, at which the Mg doping has been considered.
	Li vacancies, which
	are negatively charged and largely available in the crystal, are known to act as compensating
	defects to achieve charge neutrality in real samples. Electron polarons which are either located
	at a regular Nb lattice sites (free polarons) or at Nb substitutionals at the Li lattice site
	(bound polarons) \cite{Schirmer2009,Krampf2021}, are not explicitly considered in the models.}

\CR{The most favorable position for the Mg$^{2+}$ ions is at about 4\,{\AA} from the DW. At this
	distance, the substitutionals have a formation energy which is about 0.05\,eV lower than the
	corresponding value in the crystal bulk, suggesting increased defect concentration at the DWs.
	Exactly as in the case of bulk LN, Mg$^{2+}_{\text{Li}}$ defects at the DW only have a minor
	effect on the crystal structure.}

\begin{figure}[htbp]
	\centering
	\includegraphics[width=8cm]{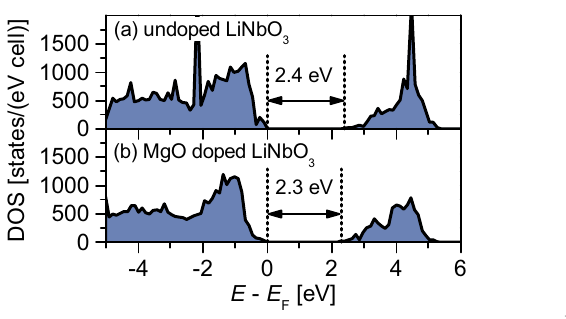}
	\caption{\CR{DFT-calculated total DOS of a supercell modelling a X-DW in the stoichiometric, 
			undoped LN crystal (panel (a)) and in the MgO-doped crystal (panel (b)).
			The focus is set on the bandgap region. The onset of the energy scale is set to the valence band edge of the undoped structure.}}
	\label{DOS_Mg:LN}
\end{figure}

\CR{Fig.\ref{DOS_Mg:LN} represents a comparison of the DOS calculated for the supercell modelling a X-DW in pure (undoped) and MgO-doped LN (panel (a) and (b), respectively). The image shows  the the DOS calculated for the most stable Mg\textsuperscript{2+} configuration and a concentration of \SI{2}{\mole\percent}. The DOS calculated  for defects at a different distance from the DW does not substantially differ. The calculations reveal that Mg ions do not significantly alter the electronic structure around the DWs, independently from their distance from the latter. Thus, insulating segments of the DWs will remain insulating also in presence of Mg ions, which are not directly involved in the conductivity enhancement. In turn, this confirms that calculations for the undoped material are already a reasonable representation of the Mg doped samples.}

\CR{In the light of the atomistic models, the possible mechanisms that lead to the formation of
	conducting DWs in MgO-doped LN, summarized in Fig.~\ref{Bandstruktur}, are evaluated:}

\begin{itemize}
	\item \CR{Part (a) shows H2H and T2T DWs becoming charged due to strong band bending induced
		by the spontaneous polarization. The latter distorts the band structure so that conduction
		states are lowered below the Fermi energy at the H2H wall and valence states are raised above
		the Fermi energy at the T2T wall. This leads to the accumulation of charge carriers (electrons
		or holes) at the DW, i.e., to charged DWs. Our calculations confirm that in LN the spontaneous
		polarization is strong enough to bend the electronic states and form localized charge carrier
		accumulation (see Fig. \ref{fig:4}).}
	
	\item  \CR{Part (b) shows a DW becoming conducting due to a local rearrangement of the bulk band
		structure as induced by the DW formation. This might lead to a substantial lowering of the
		bandgap, eventually resulting in the crossing of the Fermi energy by the conduction band (or
		valence band). Our calculations show that indeed X-DWs and Y-DWs are characterized by a fundamental
		bandgap about \SI{1.0}{\electronvolt} and \SI{0.5}{\electronvolt} smaller than the corresponding bulk value. This bandgap reduction is, however, by far not sufficient to render X-DWs and Y-DWs conducting.}
	
	\item \CR{Part (c) shows a DW that becomes conducting due to the accumulation of defects in its
		neighborhood. The defects might either introduce electronic states at the Fermi energy (electron
		conductivity) or, in case of charged defects with high mobility, directly transport charges by
		migration and diffusion (ionic conductivity). Although our calculations reveal somewhat lower
		defect formation energy of Mg-substitutionals at the DWs, they rule out the formation of Mg-related
		conducting defect bands within the fundamental bandgap.}
\end{itemize}

\CR{Concluding, ideal, straight H2H and T2T DWs in LN are characterized by such a large band bending,  that the conduction bands cross the Fermi energy at the H2H DWs and the valence bands cross the  Fermi energy at the T2T DWs as indicated in the scheme of Fig.~\ref{Bandstruktur}. This leads to  the formation of charged regions. However, as revealed by the DOS, the 180° DWs (called X-DWs  and Y-DWs in the present work) merely reduce the bandgap of the bulk crystal, which remains  insulating and not charged (see Fig.~\ref{DOS_Mg:LN}). This situation is only slightly altered by the  presence of Mg-defects. The effects on the band structure are so minor that they can be largely neglected in terms of the transport processes. Thus, the measured, thermally activated conductivity must be related to  hopping processes of other defects such as free and bound polarons.}

\section{\label{sec:summary}Conclusions and outlook}

Two measurement approaches were used to characterize the \DWC which are direct current (DC) and alternating current (AC) measurements. The results agree very well. AC measurements provide more detailed information
and enable, for example, the identification of artifacts. It can be regarded as an initial method for \DWC determination. Provided that no artifacts occur, DC measurements are suitable for routinely characterization.

Our results consistently demonstrate that as temperature rises, there is a corresponding increase in the \DWC. It is shown up to about \SI{400}{\degreeCelsius}, marking the highest temperatures documented in
the existing literature. This indicates the potential for developing DW-based sensors and electronics for medium and high-temperature applications.

At temperatures up to about \SI{400}{\degreeCelsius}, the thermally-activated nature of the \DWC is demonstrated, confirming that the inclined DW sections govern the overall behavior. Thereby, the DW current rises by several orders of magnitude with the temperature being significantly higher than the bulk LN current over the entire temperature range investigated. At e.g. \SI{400}{\degreeCelsius} the \DWC is about 4.5 orders of magnitude higher compared to LN volume crystals without additional DWs.

The measured activation energies are compatible with polaron hopping, indicating that a hopping process (or a similar mechanism) dominates the overall current. Furthermore, DWs display distinct activation energies in two overlapping temperature regimes, hinting towards the involvement of more than one type of defect/polaron to conduction, consistently observed within several DWs.

First-principles calculations demonstrate the feasibility to describe DW band structures that enable high DW currents. However, the conductivity of inclined domain walls cannot be solely explained by the formation of a 2D carrier gas and must be supported by hopping processes. This holds true up to at least \SI{400}{\degreeCelsius}.
\CR{Another essential finding from DFT calculations is that MgO doping causes only slight changes in the density of states in the bandgap region. Since MgO doping hardly affects the width of the fundamental bandgap and does not introduce any localized defect states, it can be concluded that MgO doping has a negligible direct influence in the DW current.}

\newpage

\section{Acknowledgments}
	The authors gratefully acknowledge financial support by the Deutsche Forschungsgemeinschaft (DFG) through the Research unit FOR5044 (ID: 426703838; \linebreak
	\mbox{ \url{https://www.for5044.de}}; subprojects: \mbox{FR1301/42}, \mbox{SA1948/3}, \mbox{EN434/49}, \mbox{RU2474/1}). 
	We thank Thomas Gemming and Dina Bieberstein for assistance in wafer dicing, as well as Henrik Beccard for assistance in sample preparation. EB and LME thank the Würzburg-Dresden Cluster of Excellence on “Complexity and Topology in Quantum Matter” - ct.qmat (EXC 2147; ID 39085490). Calculations for this research were conducted on the Lichtenberg high-performance computer of the TU Darmstadt and at the H\"ochstleistungrechenzentrum Stuttgart (HLRS). The authors furthermore acknowledge the computational resources provided by the HPC Core Facility and the HRZ of the Justus-Liebig-Universit\"at Gie{\ss}en.

\vspace{3cm}

\section{Data Availability Statement}
The data that support the findings of this study are available from the corresponding author upon reasonable request.

\clearpage



\appendix

\section*{Appendix}

\vspace*{0.1cm}

\section{Electrode deposition}
\label{app_electrodes}

Individual domains are electrically contacted by depositing \SI{10}{\nano\meter} thick chromium electrodes via magnetron sputtering onto both z-facets of the LN crystal. These electrodes fully cover the poled region and maintain optical transparency. Copper wires are attached to the chromium electrodes using silver paste for electrical connection with our measuring units.  	
These were used for inverting the domains and their enhancement as well as for reference measurements and characterization at room temperature. 
During the procedure, the electrode on the z$^-$ side is grounded, while the varying voltage is applied at the electrode on the z$^+$ side through a Keithley 6517B electrometer, enabling macroscopic measurement of voltage and current through the domain wall.

For the investigation of the DWs at high temperatures, approximately \SIrange{200}{300}{\nano\meter} thick Pt\textsubscript{90}Rh\textsubscript{10} electrodes were applied directly above or below the inverted domain using pulsed laser ablation (KrF excimer laser, Lambda-Physics COMPex 205, Germany). Their diameter is \SI{3}{\milli\meter}. It should be noted that initially approx. \SI{5}{\nano\meter} thick Ti-adhesive layers were deposited. The deposition parameters for Pt\textsubscript{90}Rh\textsubscript{10} (Ti) show a pulse energy of \SI{300}{\milli\joule} (\SI{200}{\milli\joule}) with a deposition time of \SI{90}{\min} (\SI{2.5}{\min}). The base pressure before deposition was \SI{1E-4}{\pascal}.

\newpage

\vspace*{0.0001cm}

\section{Domain wall preparation procedure}
\label{app_DW-preparation}

Domain walls are created by irradiating the samples locally with a laser ($\uplambda$~=~\SI{325}{\nano\meter}, P = \SI{10}{\micro\watt}) and applying a voltage of \SI{0.8}{\kilo\volt} between the surfaces of the sample using liquid electrolyte electrodes, which forces domain formation in the opposite direction to the original monodomain.
The inverted domains prepared in this way are then enhanced, see \ref{app_enhancing}.

\vspace{2.5cm}



\section{Domain images and current-voltage characteristics of all investigated domains}
\label{app_domains}

\begin{figure*}[htbp]
	\centering
	\begin{subfigure}{0.15\textwidth}
		\centering
		\includegraphics[height=3.3cm]{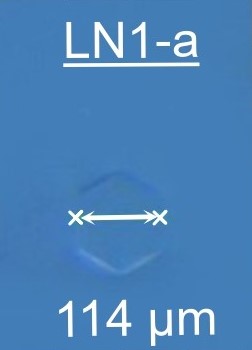}
		\caption{} 
		\label{fig:subfig_a1}
	\end{subfigure}
	\hfill
	\begin{subfigure}{0.40\textwidth}
		\centering
		\includegraphics[height=3.3cm]{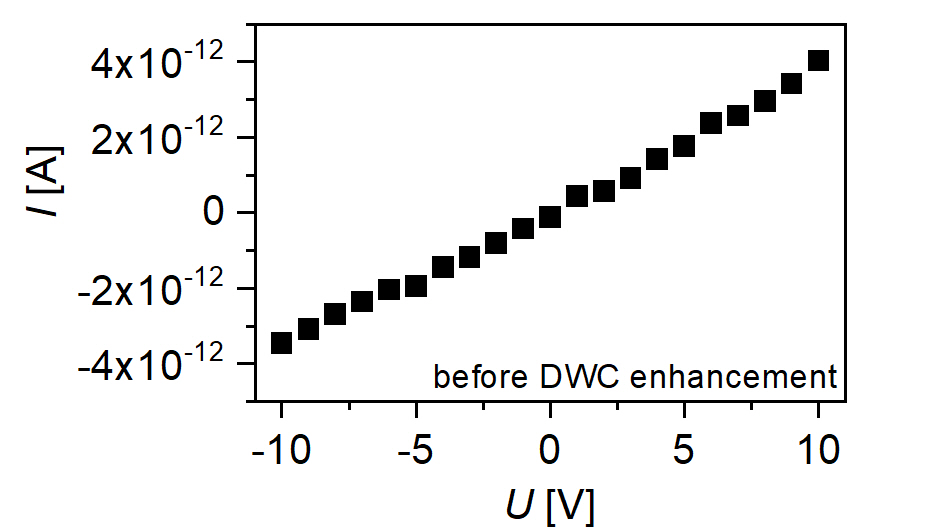}
		\caption{}
		\label{fig:subfig_b1}
	\end{subfigure}
	\hfill
	\begin{subfigure}{0.40\textwidth}
		\centering
		\includegraphics[height=3.3cm]{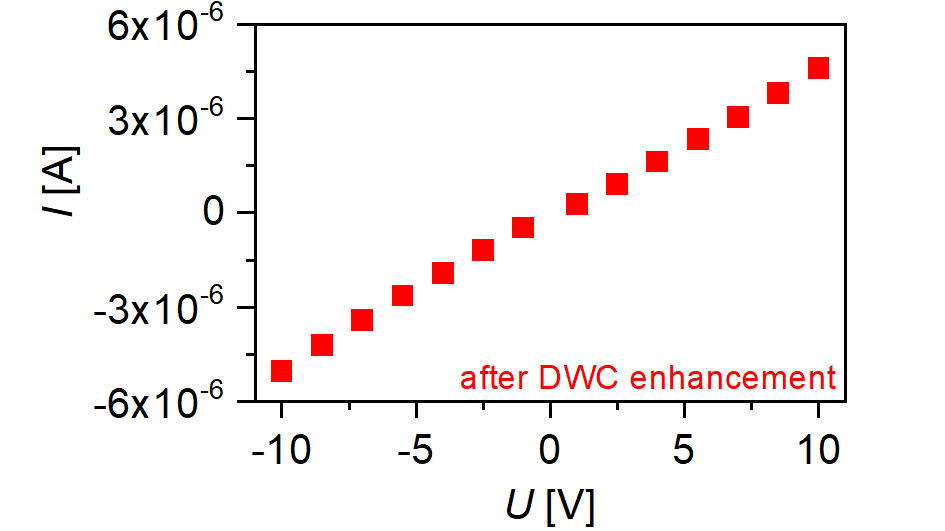}
		\caption{}
		\label{fig:subfig_c1}
	\end{subfigure}
	\captionsetup{width=16cm}
	\caption{\textbf{LN2-a:} (a) Domain image; (b) current-voltage characteristics after poling,
		and (c) after \DWC enhancement.} 
	\label{fig:A2} 
\end{figure*}

\begin{figure*}[htbp]
	\centering
	\begin{subfigure}{0.15\textwidth}
		\centering
		\includegraphics[height=3.3cm]{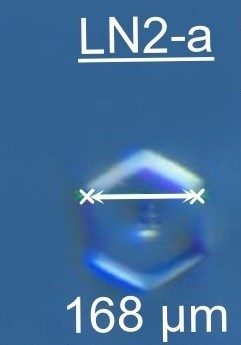}
		\caption{} 
		\label{fig:subfig_a2}
	\end{subfigure}
	\hfill
	\begin{subfigure}{0.40\textwidth}
		\centering
		\includegraphics[height=3.3cm]{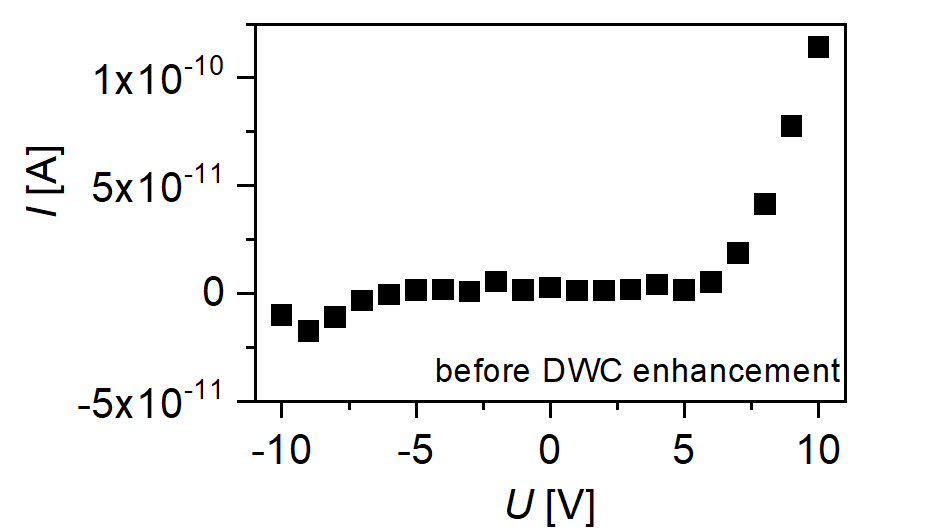}
		\caption{}
		\label{fig:subfig_b2}
	\end{subfigure}
	\hfill
	\begin{subfigure}{0.40\textwidth}
		\centering
		\includegraphics[height=3.3cm]{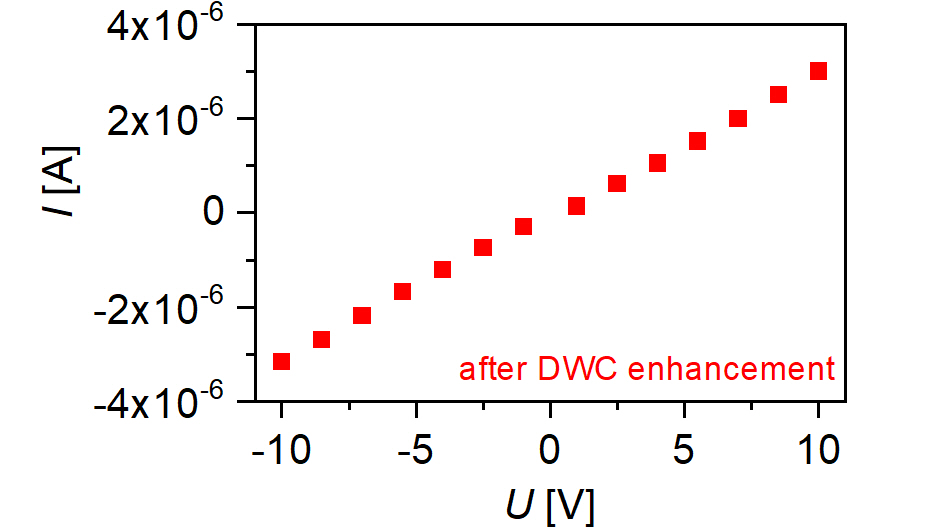}
		\caption{}
		\label{fig:subfig_c2}
	\end{subfigure}
	\captionsetup{width=16cm}
	\caption{\textbf{LN2-a:} (a) Domain image; (b) current-voltage characteristics after poling,
		and (c) after \DWC enhancement.} 
	\label{fig:A2} 
\end{figure*}

\begin{figure*}[htbp]
	\centering
	\begin{subfigure}{0.15\textwidth}
		\centering
		\includegraphics[height=3.3cm]{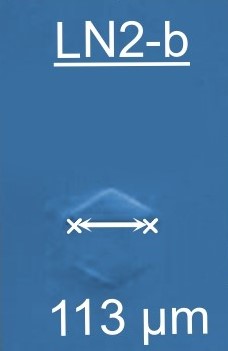}
		\caption{} 
		\label{fig:subfig_a3}
	\end{subfigure}
	\hfill
	\begin{subfigure}{0.40\textwidth}
		\centering
		\includegraphics[height=3.3cm]{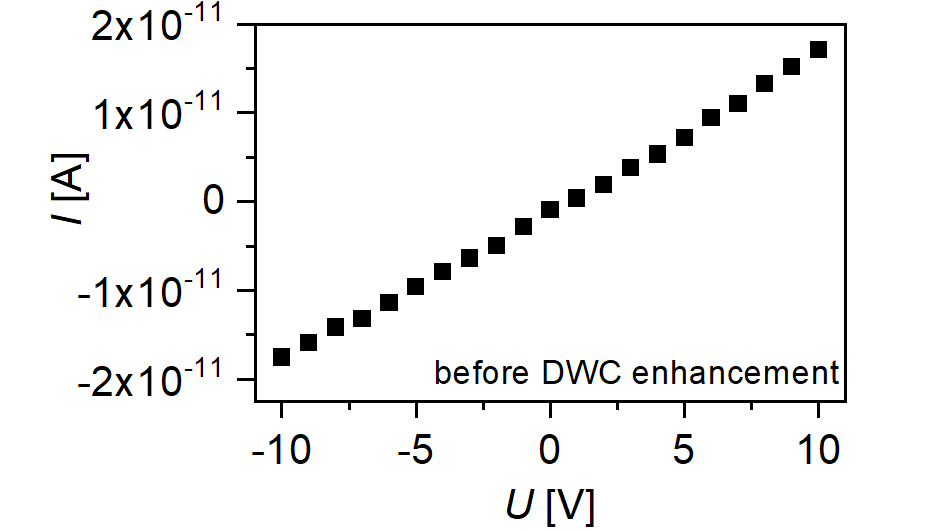}
		\caption{}
		\label{fig:subfig_b3}
	\end{subfigure}
	\hfill
	\begin{subfigure}{0.40\textwidth}
		\centering
		\includegraphics[height=3.3cm]{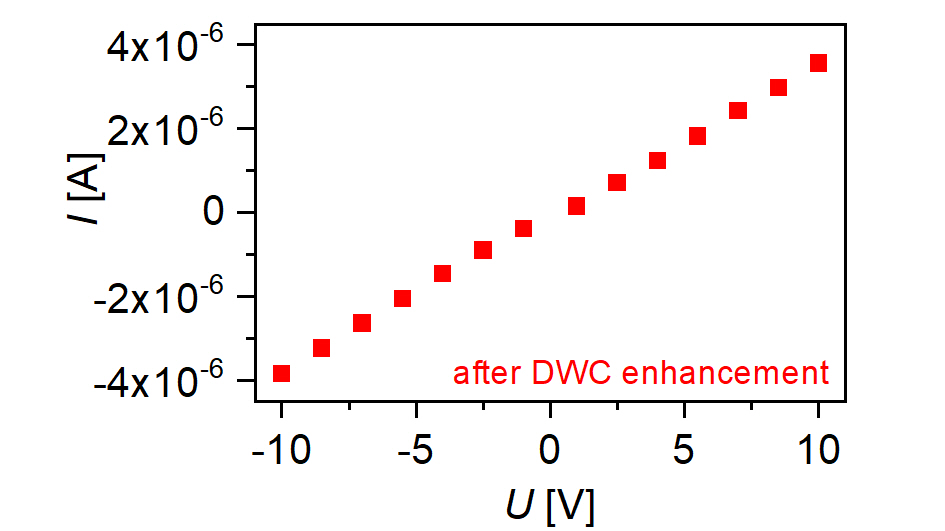}
		\caption{}
		\label{fig:subfig_c3}
	\end{subfigure}
	\captionsetup{width=16cm}
	\caption{\textbf{LN2-b:} (a) Domain image; (b) current-voltage characteristics after poling,
		and (c) after \DWC enhancement.} 
	\label{fig:A3} 
\end{figure*}

\begin{figure*}[htbp]
	\centering
	\begin{subfigure}{0.15\textwidth}
		\centering
		\includegraphics[height=3.3cm]{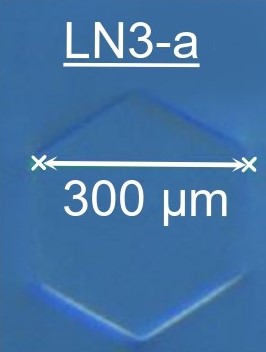}
		\caption{} 
		\label{fig:subfig_a}
	\end{subfigure}
	\hfill
	\begin{subfigure}{0.40\textwidth}
		\centering
		\includegraphics[height=3.3cm]{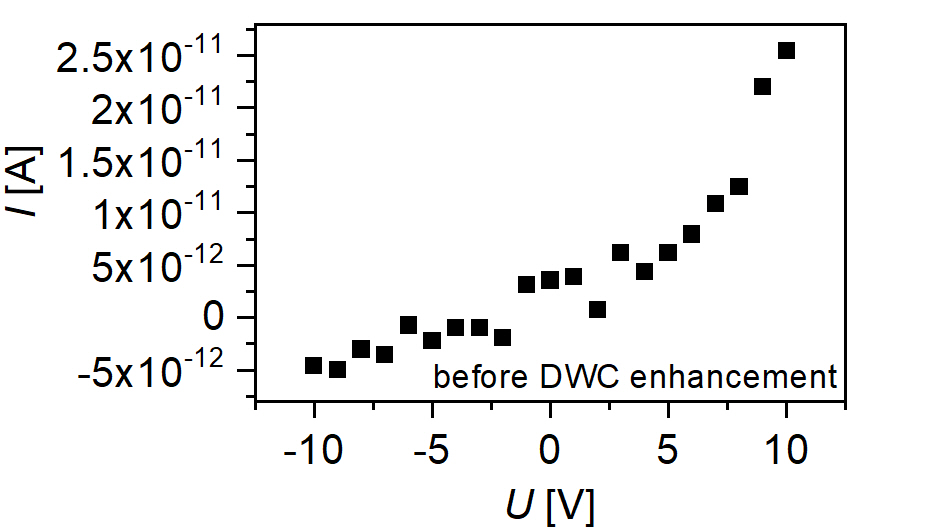}
		\caption{}
		\label{fig:subfig_b}
	\end{subfigure}
	\hfill
	\begin{subfigure}{0.40\textwidth}
		\centering
		\includegraphics[height=3.3cm]{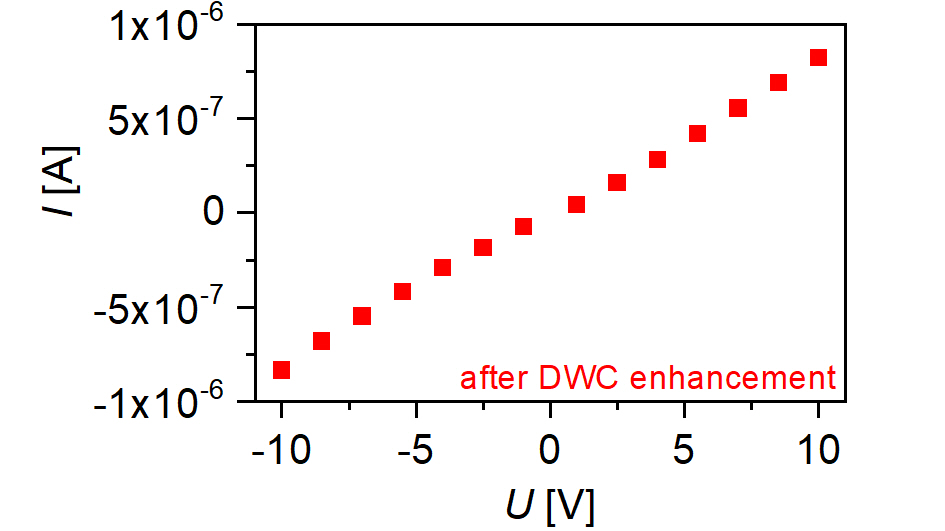}
		\caption{}
		\label{fig:subfig_c}
	\end{subfigure}
	\captionsetup{width=16cm}
	\caption{\textbf{LN3-a:} (a) Domain image; (b) current-voltage characteristics after poling,
		and (c) after \DWC enhancement.} 
	\label{fig:4-1} 
\end{figure*}

Following the UV-assisted poling process, images of the initial hexagonal domains are captured using a polarizing microscope. The domain images for LN1-a, LN2-a, LN2-b and LN3-a taken after
poling, are shown in Figs. \ref{fig:subfig_a1}, \ref{fig:subfig_a2}, \ref{fig:subfig_a3}, and \ref{fig:subfig_a}. The current-voltage (\IV) characteristics of these domain walls (DWs) are measured using an electrometer, and presented in Figs. \ref{fig:subfig_b1}, \ref{fig:subfig_b2}, \ref{fig:subfig_b3}, and \ref{fig:subfig_b}. After domain wall conductivity (DWC) enhancement, the \IV\ characteristics shown in Figs. \ref{fig:subfig_c1}, \ref{fig:subfig_c2}, \ref{fig:subfig_c3}, and \ref{fig:subfig_c} result.
After poling, \IV\ characteristics of domain LN1-a (Fig. \ref{fig:subfig_b1}) and LN2-b (Fig. \ref{fig:subfig_b3}) exhibit a linear relationship, thereby indicating ohmic behavior. However, after poling, the \IV\ curve obtained for domains LN2-a (Fig. \ref{fig:subfig_b2}) and LN3-a (Fig. \ref{fig:subfig_b}) exhibits a non-linear behavior pointing to diode-like contributions. All experiments presented in \ref{app_domains} are conducted at room temperature.


\clearpage

\section{Procedure for enhancing domain wall conductivity}
\label{app_enhancing}

The procedure for enhancing the domain wall current follows the methodology outlined by Godau et al \cite{Godau2017}. 
Initially, as-grown domain walls exhibit low current at room temperature. However, employing the specified protocol resulted in a significant enhancement of the current by several orders of magnitude. 
The electric field is increased linearly at a rate of \SI{2.5}{\volt\per\milli\meter} per second, reaching maximum values of $+$\SI{4.0}{\kilo\volt\per\milli\meter} for positive voltages and \SI{-3.6}{\kilo\volt\per\milli\meter} for negative voltages. This process results in a substantial increase in the current through of the domain walls, by up to six orders of magnitude, as demonstrated in the \IV\ characteristics of LN1-a, LN2-a, LN2-b and LN3-a domains (Figs. \ref{fig:subfig_c1}, \ref{fig:subfig_c2}, \ref{fig:subfig_c3}, and \ref{fig:subfig_c}).

\CR{Note that a DC bias up to $\pm$\SI{10}{\volt} as applied during the electrometer measurements and impedance spectroscopy is about two orders of magnitude lower than those applied by during sample preparation (e.g. domain modification at \SI{800}{\volt}). Therefore, it is very unlikely that the voltages of up to $\pm$\SI{10}{\volt} will change or degrade the samples. In \cite{Kirbus2019}, DWs were investigated by second-harmonic generation (SHG) while applying high voltages (up to several kVs) giving no evidence for degradational effects for the comparibly low electrical fields used in this research. 
	For nanoscale devices such voltages are critical, but since this work is about \SI{200}{\micro\meter} thick crystals, the voltage drop along the DW into the depth is at the maximum of $\pm$\SI{10}{\volt} still only \SI{50}{\volt\per\milli\meter} and even lower for lower bias voltages and with this ways off from any depolarizing field.}

\newpage

\section{Micro-impedance setup and measurement uncertainty}
\label{app_micro-impedance-setup}

\begin{figure}[H]
	\centering
	\includegraphics[scale=0.235]{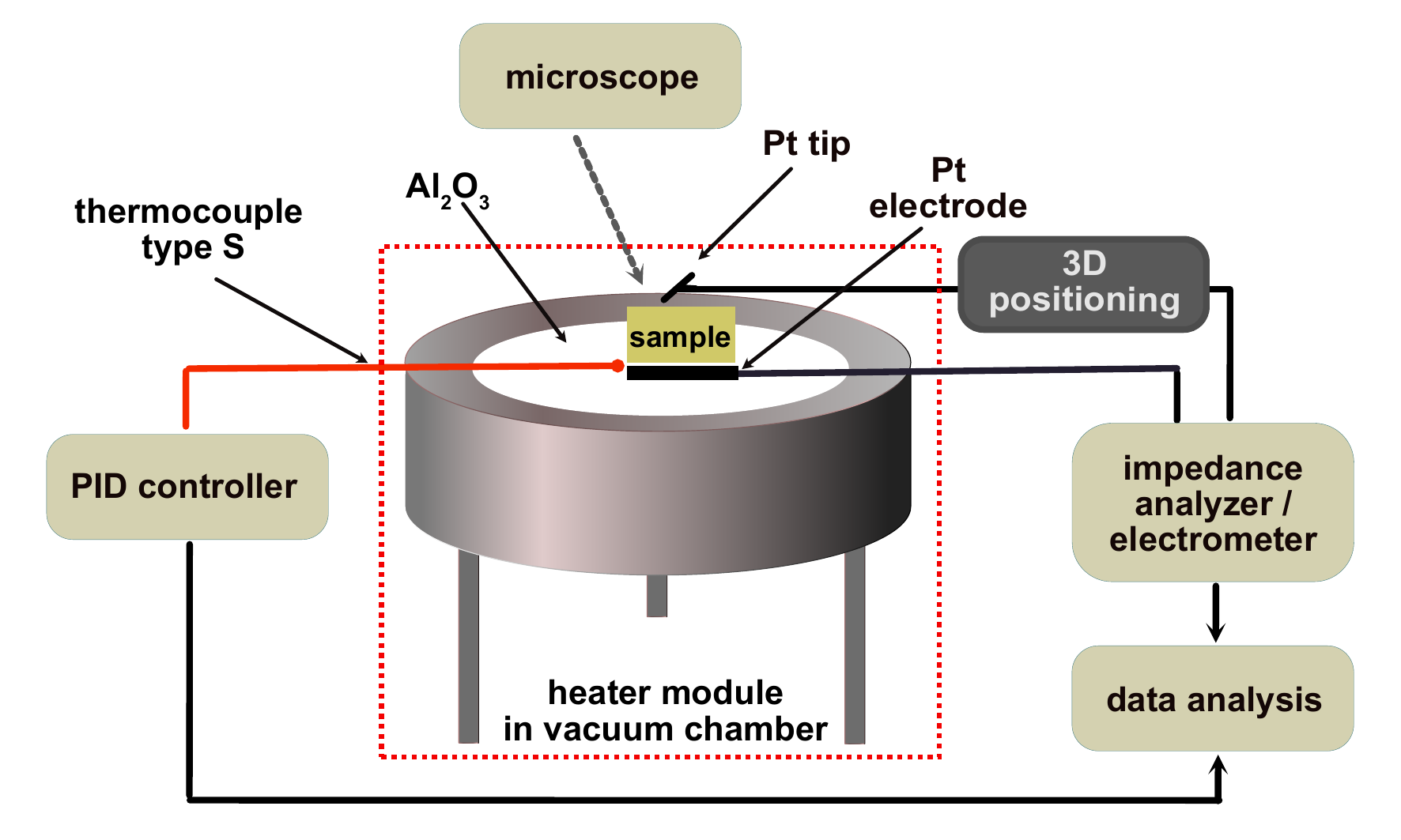}
	\caption{Micro-impedance setup for high temperature measurement.}
	\label{fig:A5}
\end{figure}

To measure DW currents at high temperatures, a micro-impedance setup is employed. This setup includes a heater module (UHV Design, UK) and precise positioning screws for the top-contact platinum tip, as depicted in Fig. \ref{fig:A5}. In this setup, the sample lays on a Pt electrode, both of which are placed on a ceramic plate atop the heater module. The heating element of the module is made of high-density graphite coated with pyrolytic graphite. The entire setup, including the sample, electrodes, and heater module, is enclosed in a vacuum chamber maintained at a total pressure of approximately \SI{2}{\pascal}. Inside the chamber, a type S thermocouple is places within the ceramic plate on top of the heater module with a distance of about \SIrange{1}{2}{\milli\meter} from the sample.

This chamber is connected to either a highly sensitive amperometer or an impedance analyzer, see section \ref{experimental_setup}. \IV\,~sweeps ranging from \SI{-10}{\volt} to $+$\SI{10}{\volt} are applied, with a step size of \SI{1.5}{\volt}. Each sweep takes around \SI{14}{\second} to complete. This method enables data collection at different temperatures, with a heating rate of \SI{1}{\kelvin\per\min}. The setup reaches a maximum temperature of \SI{400}{\degreeCelsius}.

The uncertainties of the DC measurement for currents of \SI{0.01}{\micro\ampere}, \SI{0.1}{\micro\ampere} and \SI{1}{\micro\ampere} are $\pm$\SI{0.51}{\nano\ampere}, $\pm$\SI{6}{\nano\ampere}
and $\pm$\SI{1.5}{\nano\ampere} respectively (range: \SI{20}{\micro\ampere}). The same applies to the voltage accuracy. In the most unfavorable case of $\vert U\textsubscript{M}\vert = \SI{10}{\volt}$ for the measured characteristics, this corresponds to a relative uncertainty of \SI{0.25}{\percent} or $\pm$\SI{25}{\milli\volt}.


\section{Temperature dependent \textit{IV} curves}
\label{app5}

\begin{figure}[htbp]
	\centering
	\includegraphics[scale=1.0]{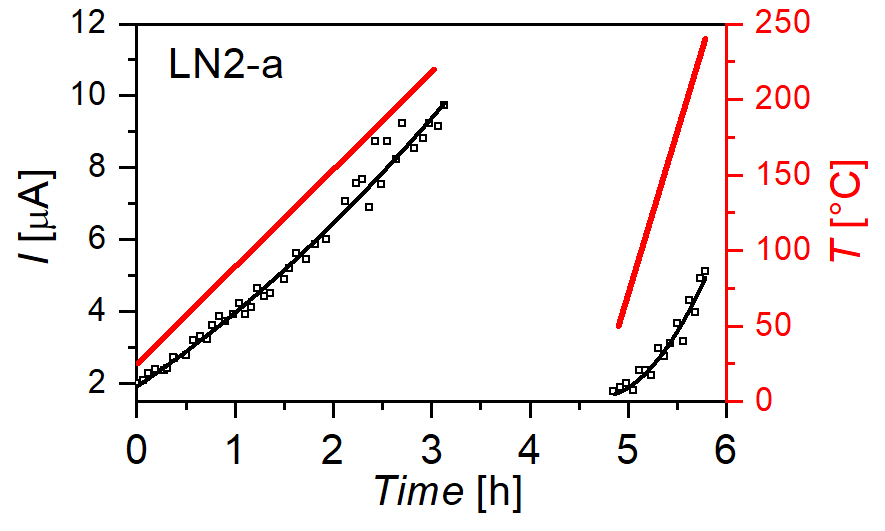}
	\caption{Time-dependent temperature and DW current at a given voltage of $+$\SI{7}{\volt}.}
	\label{fig:A6}
\end{figure}

Following the enhancement of \DWC, Fig.~\ref{fig:A6} illustrates the temperature and \DWC of LN2-a as function of time at a given potential of $+$\SI{7}{\volt} as an example. In the case of the current dependency, the data is represented by a scatter plot with open symbols, while the line serves only as a guide for the eyes. The example LN2-a is used to demonstrate the repeated heating over time to illustrate the experimental procedure. As shown, the initial heating cycle is done up to \SI{200}{\degreeCelsius}. Subsequent heating cycle was carried out at higher temperature. 

\begin{figure}[H]
	\centering
	\begin{subfigure}{0.49\textwidth}
		\centering
		\includegraphics[scale=1]{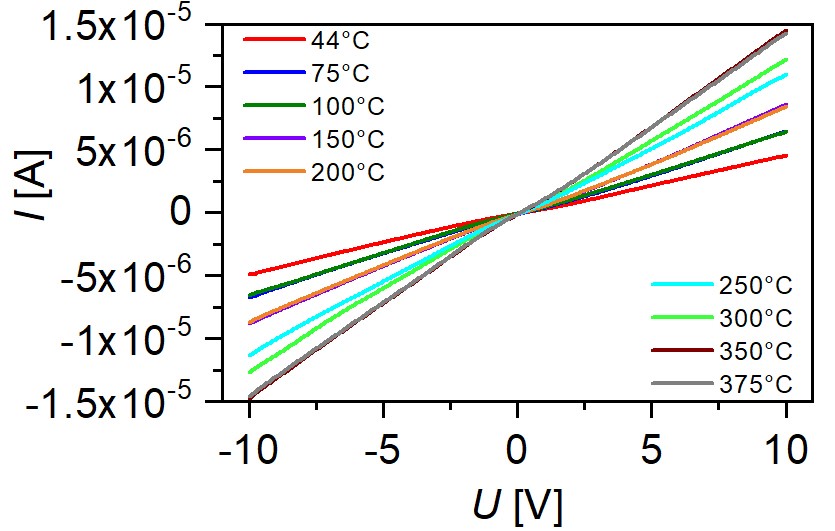}
		\caption{\textbf{LN1-a}}
		\label{fig:subfig_a4}
	\end{subfigure}
	\hfill
	\begin{subfigure}{0.49\textwidth}
		\centering
		\includegraphics[scale=1]{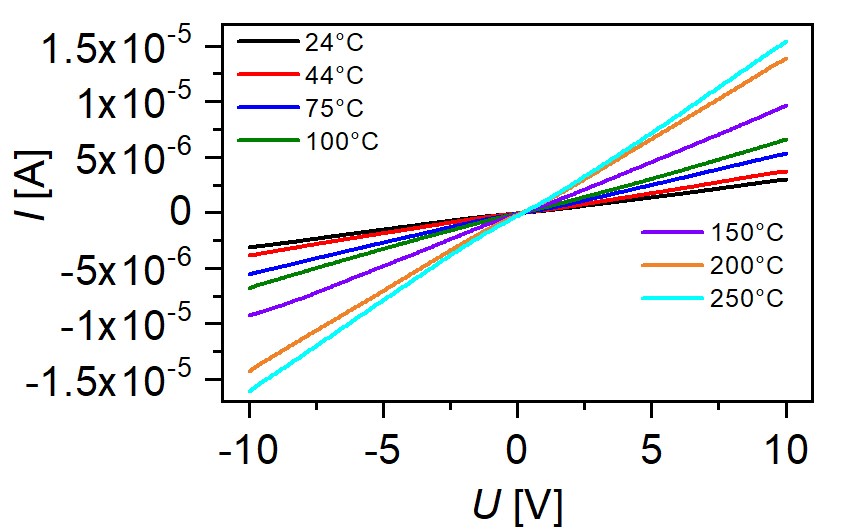}
		\caption{\textbf{LN2-a}}
		\label{fig:subfig_b4}
	\end{subfigure}
	\\
	\begin{subfigure}{0.49\textwidth}
		\centering
		\includegraphics[scale=1]{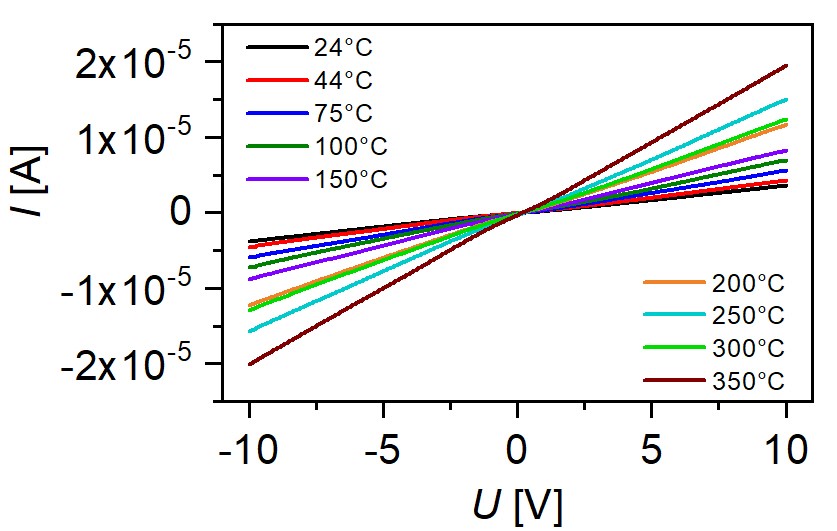}
		\caption{\textbf{LN2-b}} 
		\label{fig:subfig_c4}
	\end{subfigure}
	\caption{\IV\ curve across various temperatures, up to \SI{375}{\degreeCelsius}. Note the ohmic behavior of all \IV\ curves, being independent on temperature or domain size.} 
	\label{fig:A7} 
\end{figure}

As shown in section \ref{T-depIV} (Fig.~\ref{fig:1}) for domain LN3-a, the \IV\ curves of domains LN1-a, LN2-a, and LN2-b also exhibit ohmic behavior. Figs. \ref{fig:subfig_a4}, \ref{fig:subfig_b4}, and \ref{fig:subfig_c4} illustrate the \IV\ curves at different temperatures, reaching up to \SI{375}{\degreeCelsius}. 

\newpage

\vspace*{0.7cm}

\newpage

\section{Temperature-dependent domain wall current in LN2-a and \mbox{LN2-b} and activation energies of the \DWC }
\label{app6}

\begin{figure}[htbp]
	\centering
	\includegraphics[scale=0.99]{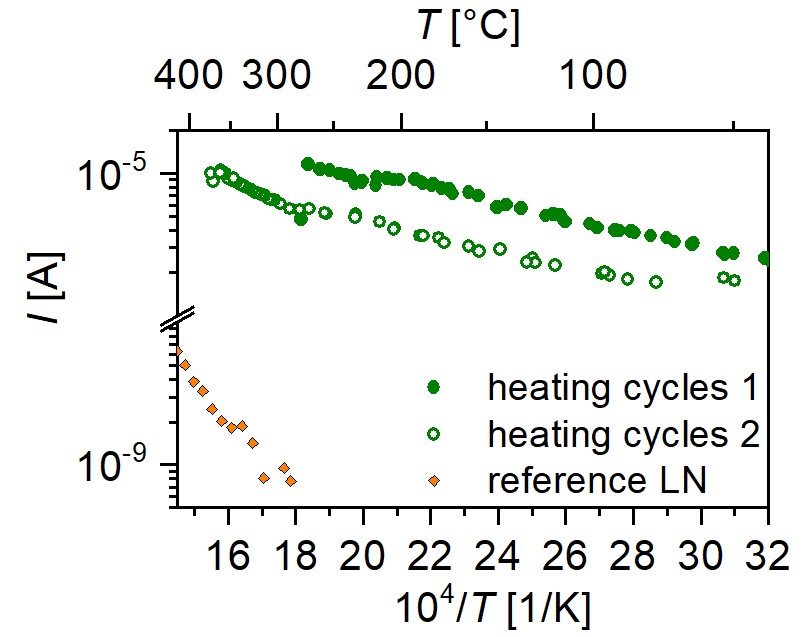}
	\caption{Temperature-dependent \DWC of LN2-a at a given voltage of $+$\SI{7}{\volt} for for two subsequent heating cycles. }
	\label{fig:A8}
\end{figure}

\begin{figure}[htbp]
	\centering
	\includegraphics[scale=0.99]{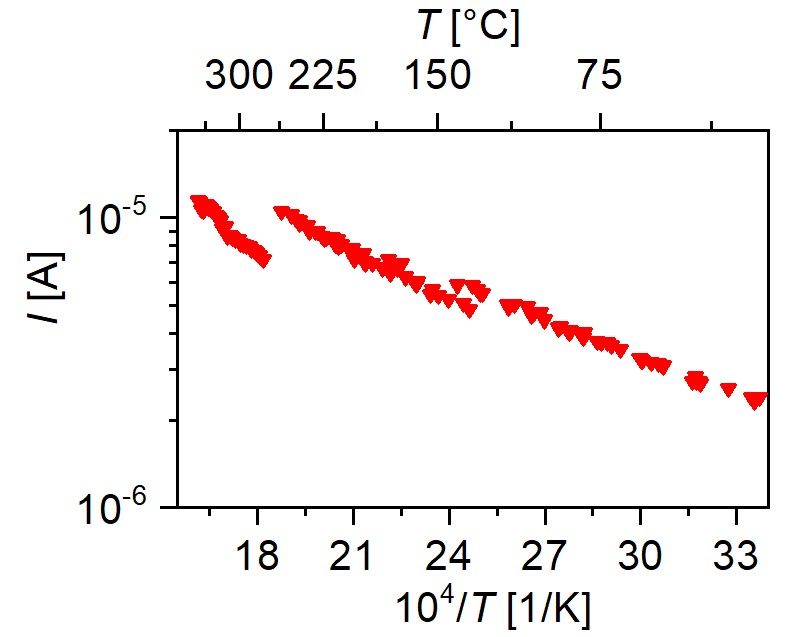}
	\caption{Temperature-dependent \DWC of LN2-b at a given voltage of $+$\SI{7}{\volt}. }
	\label{fig:A9}
\end{figure}

\begin{figure}[htbp]
	\centering
	\begin{subfigure}{0.48\textwidth}
		\centering
		\includegraphics[scale=0.9]{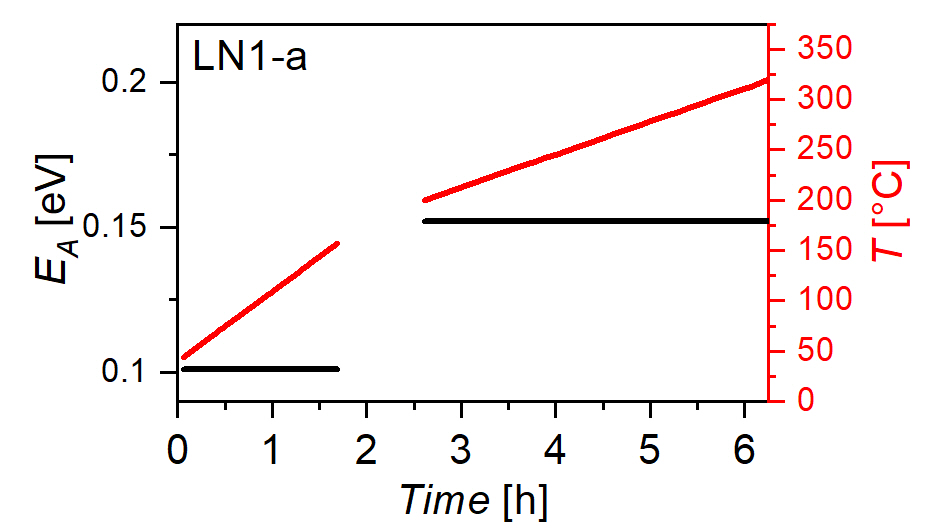}
		\caption{\textbf{ }}
		\label{fig:subfig_a5}
	\end{subfigure}
	\hfill
	\begin{subfigure}{0.48\textwidth}
		\centering
		\includegraphics[scale=0.9]{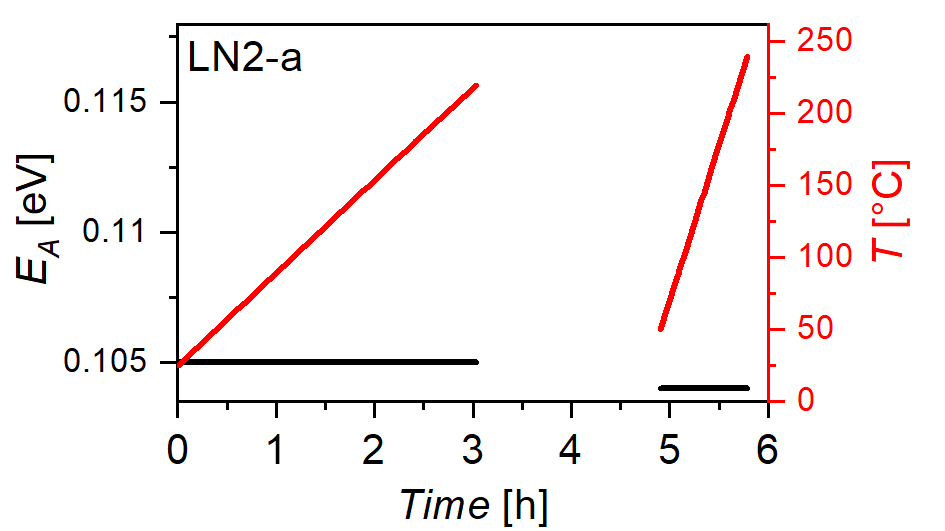}
		\caption{\textbf{ }}
		\label{fig:subfig_b5}
	\end{subfigure}
	\\
	\begin{subfigure}{0.48\textwidth}
		\centering
		\includegraphics[scale=0.9]{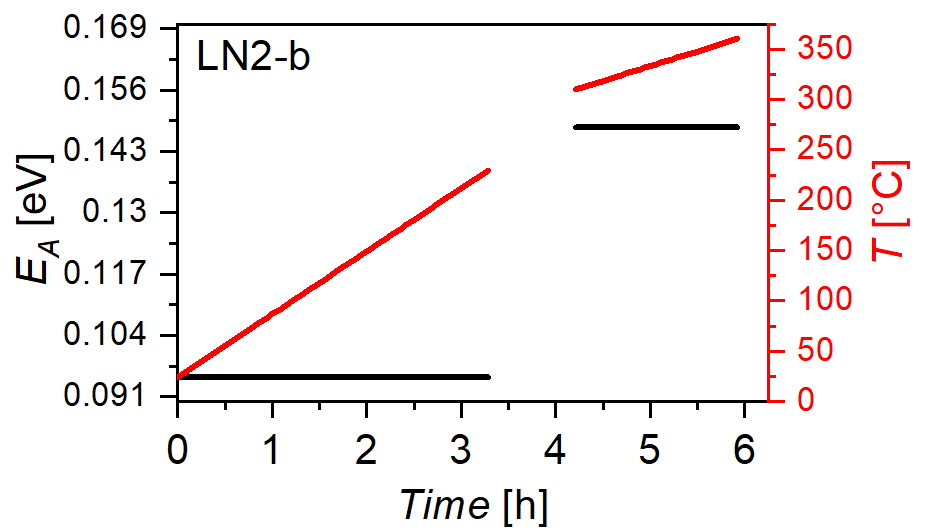}
		\caption{\textbf{ }} 
		\label{fig:subfig_c5}
	\end{subfigure}
	\hfill
	\begin{subfigure}{0.48\textwidth}
		\centering
		\includegraphics[scale=0.9]{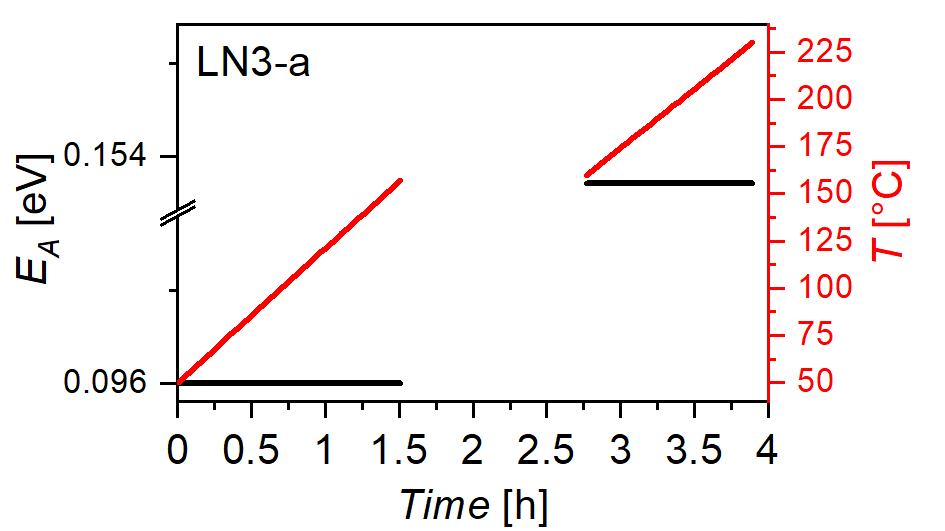}
		\caption{\textbf{ }}
		\label{fig:subfig_d5}
	\end{subfigure}
	\caption{Time-dependent temperature and activation energies of the \DWC at $+$\SI{7}{\volt}. } 
	\label{fig:A10} 
\end{figure}

Figures \ref{fig:A8} and \ref{fig:A9} depict the \DWC measured at a bias voltage of $+$\SI{7}{\volt} for LN2-a and LN2-b, respectively, presented in Arrhenius-type plots. Notably, both figures illustrate a nearly linear increase in \DWC in this representation, although with varying slopes. This observation also suggests the presence of thermally-activated transport processes characterized by different activation energies. For better visualization, Fig.~\ref{fig:A10} provides the activation energies for different temperature ranges as funtion of time.The data shown in Fig.~\ref{fig:A10} represent results only for the linear regions in the Arrhenius plot. For the non-linear regions, we cannot accurately determine the activation energy, and therefore these data are not presented, resulting in gaps in the figure.

\newpage

\section{Atomistic calculations}
\label{app_atomistic}

\begin{figure}[htbp]
	\centering
	\includegraphics[scale=0.7]{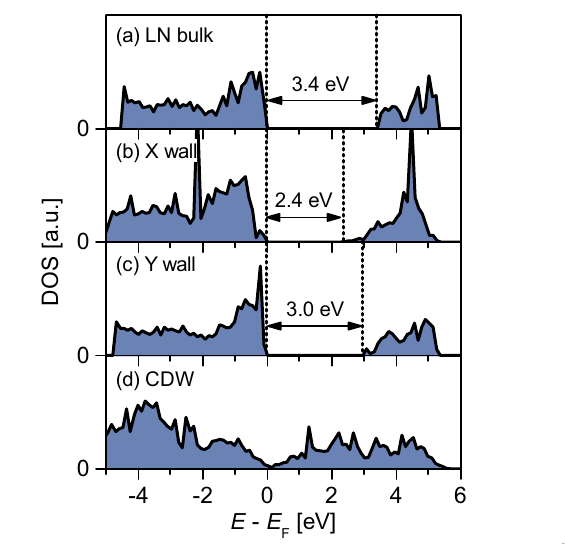}
	\caption{ DFT-calculated total DOS of LN for the crystal bulk (a) and different wall types: (b) X walls, (c) Y walls, (d) H2H and T2T walls, labeled as charged domain walls (CDW).}
	\label{fig:11}
\end{figure}

Atomistic calculations as described in the main article were performed on supercells of different symmetries. X and Y DWs are parallel to the crystal X and Y axis, and thus parallel to the (10$\overline{1}$0) and (11$\overline{2}$0) LN lattice planes, respectively. Both types of DWs can be modelled by means of orthorhombic supercells. A 1$\times$8$\times$1 and a 6$\times$1$\times$1 repetition of the orthogonal unit cell are employed to model X and Y DWs, respectively. Within these supercells, two X and Y DWs are separated by about \SI{36}{\angstrom} and by about \SI{15}{\angstrom}. 3$\times$3$\times$3 and 4$\times$3$\times$2 Monkhorst-Pack k-point meshes are used for the energy integration of the supercells modelling X and Y DWs, respectively. 
H2H and T2T DWs are modelled by means of a 1$\times$1$\times$12 repetition of the hexagonal unit cell and a 4$\times$4$\times$1 Monkhorst-Pack k-point mesh. Tests with larger and smaller supercells have ben performed in all cases to ensure numerically converged results with respect to the DW separation.
Atomic structures are relaxed untill the Helman-Feynman forces acting on the single atoms are lower than a threshold of \SI{E-3}{\electronvolt\per\angstrom}.
Fig.~\ref{fig:11} shows the calculated density of states (DOS) of different structures. Panel (a) represents the DOS of LN buk. The fundamental gap of about \SI{3.4}{\electronvolt} is clearly visible. Panels (b) and (c) show the DOS of the supercells modelling X and Y DWs, respectively. In both cases the electronic gap shrinks to about \SI{2.4}{\electronvolt} and \SI{3.0}{\electronvolt}, respectively. Panel (d) illustrates the DOS calculated for the supercell modelling the fully charged domain walls (CDW), for which valence and conduction bands are merged.

\newpage

\twocolumn



  \bibliographystyle{elsarticle-num} 
  \bibliography{References_Revision}






\end{document}